\documentclass[twocolumn,aps,showpacs,floatfix,amsmath,amssymb,superscriptaddress,prb]{revtex4-2}
\usepackage{amsmath, amsfonts, amssymb, graphics, graphicx, epsfig, xcolor, times, indentfirst, layout, subfigure, multirow, bm, soul}
\usepackage[colorlinks=true,citecolor=blue,urlcolor=blue]{hyperref}
\usepackage{threeparttable}
\usepackage{dcolumn}
\usepackage{dsfont}
\usepackage{mathtools}
\usepackage{orcidlink}

\newcommand{\bc}{\begin{center}}
\newcommand{\ec}{\end{center}}
\newcommand{\be}{\begin{equation}}
\newcommand{\ee}{\end{equation}}
\newcommand{\ba}{\begin{array}}
\newcommand{\ea}{\end{array}}
\newcommand{\beq}{\begin{eqnarray}}
\newcommand{\eeq}{\end{eqnarray}}
\newcommand{\ket}[1]{\left| {#1}\right\rangle}
\newcommand{\bra}[1]{\left\langle {#1} \right|}
\newcommand{\skp}[2]{\langle {#1} | {#2}\rangle}
\newcommand{\e}[1]{\langle {#1}\rangle}
\newcommand{\abs}[1]{\left| {#1}\right|}
\newcommand{\trip}[3]{\langle {#1}|{#2}|{#3}\rangle}

\begin{document}

\title{Quench dynamics of the quantum XXZ chain with staggered interactions:
Exact results and simulations on digital quantum computers}
%\title{Quench dynamics of the quantum XXZ chain in the flat-band limit:
%Exact results and simulations on digital quantum computers}

\author{Ching-Tai Huang}
\affiliation{Graduate Institute of Applied Physics, National Chengchi University, Taipei 11605, Taiwan}

\author{Yu-Cheng Lin\,\orcidlink{0000-0001-6112-3723}}
\email{yc.lin@nccu.edu.tw}
\affiliation{Graduate Institute of Applied Physics, National Chengchi University, Taipei 11605, Taiwan}

\author{Ferenc Igl{\'o}i\,\orcidlink{0000-0001-6600-7713}}
\email{igloi.ferenc@wigner.hun-ren.hu}
\affiliation{Wigner Research Centre for Physics, Institute for Solid State Physics and Optics, H-1525 Budapest,  Hungary}
\affiliation{Institute of Theoretical Physics, University of Szeged, H-6720 Szeged, Hungary}

\begin{abstract}
We investigate quench dynamics in the quantum $S=1/2$ XXZ antiferromagnetic
chain with staggered and anisotropic interactions in the flat-band limit.  We
consider global quenches that interchange odd- and even-bond strengths in a
fully dimerized chain, yielding a postquench Hamiltonian with couplings only on
even links.  Due to the vanishing group velocity of excitations, the system
fails to relax, providing an ideal setting to study dynamical topological
quantum phase transitions.  We show that the $z$-component of the interactions,
controlled by the anisotropy $\Delta$, acts as a relevant term that
generates additional dynamical quantum phase transitions and induces rich
structures in the Loschmidt echo.  These transitions are symmetry-protected and
robust for quenches between distinct dimerization patterns.  
Using a Bell-basis representation, we derive closed-form results for entanglement entropies,
Loschmidt echoes for finite and infinite systems, and identify both
$\Delta$-independent and $\Delta$-dependent critical times. 
Quantum simulations on IBM Quantum processors demonstrate that key observables
can be reliably extracted on noisy intermediate-scale devices without error mitigation.
\end{abstract}

\maketitle

\section{Introduction}
\label{intro}

Over the past decades, non-equilibrium dynamics of closed quantum systems has
become an intensively studied topic, both theoretically and experimentally.
Among experimental platforms, ultracold atomic systems offer an ideal setting
for exploring such quantum dynamics~\cite{Weiss,Hofferberth,Trotzky,Meinert,Langen}.  
More recently, rapid advances in programmable quantum devices based on various types of physical qubits have
opened new avenues for probing the dynamics of quantum many-body
systems~\cite{Qsim_rev,LGT_PRXQ,Smith,Pollmann,Knolle_SR,Knolle_PRE,Chiral_dyn,XX,Knolle_E8,Knolle_Hung,Chen_flat}.

Central questions in this field concern the nature of relaxation and the
long-time behavior following a quantum quench, i.e. a sudden change in the
Hamiltonian parameters~\cite{Mitra}.  In homogeneous systems with extended eigenstates, 
local observables typically relax exponentially fast, and the
stationary state depends on whether the underlying Hamiltonian is integrable or
non-integrable. 
Non-integrable systems are expected
to thermalize, reaching a stationary state described by a Gibbs ensemble with
an effective temperature~\cite{Deutsch,Rigol_Nature,Polkovnikov_rev1,Polkovnikov_rev2}. On the other hand, integrable systems,
characterized by an extensive number of conserved quantities, typically
equilibrate to non-thermal stationary states described by a so-called
generalized Gibbs ensemble~\cite{Cazalilla,Rigol_PRA,Rigol_PRL,Fagotti,Caux,Caux_PRL2014,Prosen,Essler}.

There also exist systems that fail to relax after a quench.
Several mechanisms can lead to a lack of relaxation, including
confinement~\cite{Calabrese_nphys,Viti,Schwinger}, flat band~\cite{Flatband,Flatband_SciAdv}, 
disorder~\cite{Santoro,Disorder}, and many-body quantum scars~\cite{Scar_NatPhys,Scar}.  
Absence of relaxation is often reflected in the growth and spreading of entanglement;
one scenario is persistent and long-lived entanglement oscillations even in the infinite-size limit.

In this paper, we investigate quench dynamics of the quantum $S = 1/2$ XXZ antiferromagnetic chain with staggered and
anisotropic interactions in the flat-band limit.
We consider global quenches implemented by interchanging the odd- and even-bond strengths in a fully dimerized XXZ chain,
such that the postquench Hamiltonian contains couplings only on even links.
Due to the vanishing group velocity of its excitations, this system fails to relax after
the quench and thus provides a simple yet ideal platform for studying
dynamical topological quantum phase transitions. While flat-band
dynamics in the dimerized XX chain and related models have been explored
previously~\cite{Ueda,XX}, here we focus on the role of the anisotropy $\Delta$ in the
$z$-component of the interactions.
We show that the finite $ZZ$ interaction, controlled by the anisotropy
parameter $\Delta$, is a relevant term that generates
additional DQPTs, giving rise to rich structures in the Loschmidt echo.
These DQPTs are symmetry-protected and remain robust at short and intermediate time scales 
for quenches between Hamiltonians with distinct topological properties (i.e., different
dimerization patterns). An exception occurs in the dimerized XXX chain
($\Delta = 1$), where DQPTs are accidentally absent in the fully dimerized
limit but reemerge away from the flat-band regime.

We derive a general expression for the time-dependent states in the Bell basis
for arbitrary system sizes and obtain closed-form results for dynamical
entanglement entropies and Loschmidt echoes in both finite and infinite
systems. Building on the Bell-state representation and a transfer-matrix approach, 
we identify two types of critical times at which DQPTs occur: 
one independent of $\Delta$ and the other explicitly $\Delta$-dependent.
Our analysis further reveals nonmonotonic finite-size behavior of the
Loschmidt echo at critical times as a function of chain length, as well as a
$\Delta$-independent scaling law for chains with an odd number of dimers.
Finally, we numerically examine the robustness of these DQPTs by adding weak odd-bond couplings 
in the postquench Hamiltonian for representative values of $\Delta$, 
using the infinite time-evolving block decimation (iTEBD) method.

In addition, we perform numerical experiments on fully dimerized chains using the IBM Quantum processors. 
In one experiment, we estimate the coefficients of the initial state in the Bell basis
using the Hadamard test~\cite{Hadamard} and reconstruct the time-dependent state from these coefficients
to obtain dynamical entanglement entropies and the Loschmidt echo for various values $\Delta$.
In a second experiment, we implement quantum circuits for the time evolution  
and measure the evolved states in randomly selected Pauli bases.
By applying classical post-processing to the measurement outcomes using statistical correlations and
classical shadow techniques~~\cite{RM_PRA97,RM,RM_PRA,Shadow,RM_rev}, we estimate the second-order R\'enyi entanglement
entropy, the Loschmidt echo, and a topological order parameter, called twist order parameter~\cite{Lieb,Todo}.
The naturally Trotter-error-free implementation of decoupled time-evolution gates, combined
with the random Pauli measurement protocol, enables the extraction of reliable
results on current noisy intermediate-scale quantum devices without additional error mitigation.

The paper is organized as follows: In Sec.~\ref{sec:model}, we introduce the model and the quench protocols considered in this study.
We review the Bell basis and describe how to reconstruct the time-dependent states by working in this basis. 
We also define the entanglement entropies and the Loschmidt echo, along with their key properties.
In Sec.~\ref{sec:exact}, we present analytical derivations of the half-chain entanglement entropies and the Loschmidt echo for our model,
obtaining closed-form solutions for these observables for both finite and infinite systems. We discuss the crucial effects of
the anisotropy on DQPTs, and determine the corresponding critical times. Numerical results based on iTEBD 
for dimerized chains with weak odd-link couplings are also provided. In Sec.~\ref{sec:simulation}, 
we describe our simulation setups on IBM Quantum devices and present the results obtained using the Hadamard test and randomized Pauli measurements.
We conclude in Sec.~\ref{sec:summary} with a summary and discussion.   
Appendix~\ref{sec:S_n4} collects some explicit analytical expressions for the entanglement entropies;
Appendix~\ref{sec:Uz} presents the derivation of a shadow estimator for the twist order parameter.

\section{Model and its non-equilibrium dynamics}
\label{sec:model}

We consider global quantum quenches in the dimerized spin-$1/2$ XXZ Heisenberg chain with alternating bond strengths, defined
by the Hamiltonian of $N=2n$ spins (or $n$ dimers)
%\begin{align}
%{\cal H}(J,J',\Delta)&=\sum_{i=1}^n \left[J(S^x_{2i-1} S^x_{2i}+S^y_{2i-1} S^y_{2i}+\Delta S^z_{2i-1} S^z_{2i})\right.\nonumber\\
%&\left.+J'(S^x_{2i} S^x_{2i+1}+S^y_{2i} S^y_{2i+1}+\Delta S^z_{2i} S^z_{2i+1})\right]\;,
%\label{Hamilton}
%\end{align}
\be
     {\cal H}(J',J,\Delta) = {\cal H}_1(J',\Delta) + {\cal H}_2(J,\Delta)\,,
\ee
with
\be
    {\cal H}_1(J',\Delta) = \sum_{j=1}^n J'\left(S^x_{2j-1} S^x_{2j}+S^y_{2j-1} S^y_{2j}+\Delta S^z_{2j-1} S^z_{2j} \right) 
\ee
and 
\be
     {\cal H}_2(J,\Delta) = \sum_{i=1}^n J\left(S^x_{2j} S^x_{2j+1}+S^y_{2j} S^y_{2j+1}+\Delta S^z_{2j} S^z_{2j+1} \right)
\ee
where ${S}_j^\alpha,\;\alpha=x,y,z$ are the spin-1/2 operators at site $j$, $J$ and $J'$
are positive coupling constants on odd and even links, respectively. The parameter $\Delta$ characterizes the
anisotropy in the interaction.
For periodic boundary conditions (PBC), we have ${S}^\alpha_{2n+1}={S}^\alpha_1$.
In some cases, we consider chains with open boundary conditions (OBC), i.e.
\be
    {\cal H}_2(J,\Delta) = \sum_{j=1}^{n-1} J\left(S^x_{2j} S^x_{2j+1}+S^y_{2j} S^y_{2j+1}+\Delta S^z_{2j} S^z_{2j+1} \right)\,.  
    \label{eq:OBC}
\ee
%\begin{align}
%{\cal H}(J,J',\Delta)&=\sum_{i=1}^n \left[J(S^x_{2i-1} S^x_{2i}+S^y_{2i-1} S^y_{2i}+\Delta S^z_{2i-1} S^z_{2i})\right]\nonumber\\
%&+ \sum_{i=1}^{n-1}\left[J'(S^x_{2i} S^x_{2i+1}+S^y_{2i} S^y_{2i+1}+\Delta S^z_{2i} S^z_{2i+1})\right]\;,
%\label{eq:OBC}
%\end{align}
For $\Delta=0$, the model reduces to the XX chain, which can be mapped to
well-studied free-fermion systems~\cite{Lieb,SSH}. At $\Delta=1$, the system becomes the XXX
chain: in the uniform case ($J=J'$), it is exactly solvable via the Bethe
ansatz~\cite{Bethe}, while in the dimerized case it exhibits the well-known spin-Peierls 
transition~\cite{Pytte,Cross}.

We initialize the system at $t=0$ in the ground state $\ket{\Psi_0}$ of the fully dimerized
Hamiltonian $\mathcal{H}(J, 0, \Delta)=\mathcal{H}_1(J,\Delta)$ with $J>0$,
%$\mathcal{H}(J>0, J'=0, \Delta)$,
corresponding to a product of singlet dimers on odd links.
The state then evolves unitarily under $\mathcal{H}(0, J, \Delta)=\mathcal{H}_2(J, \Delta)$, %$\mathcal{H}(J=0, J'=J>0, \Delta)$. 
corresponding to a sudden interchange of the odd and even bond strengths.
In this setting, both the prequench and postquench Hamiltonians can be written as sums of noninteracting dimers:
\be
%{\cal H}{(J,0,\Delta)}=\sum_i^n {h}_{2i-1},\quad {\cal H}{(0,J,\Delta)}=\sum_i^n {h}_{2i}\;,
{\cal H}_1{(J,\Delta)}=\sum_j^n {h}_{2j-1},\quad {\cal H}_2{(J,\Delta)}=\sum_j^n {h}_{2j}\;,
\label{H1}
\ee
with
\be
h_{j}=J(S^x_{j} S^x_{j+1}+S^y_{j} S^y_{j+1}+\Delta S^z_{j} S^z_{j+1})\;.
\ee
Each dimer has eigenstates given by the Bell states, which can be expressed in the $S^z$-basis ($Z$-basis), 
$| 0 \rangle$ and $| 1\rangle$, as
\begin{align}
&|\psi_0(j)\rangle=\frac{1}{\sqrt{2}}(|0_j 1_{j+1}\rangle - | 1_j 0_{j+1}\rangle),\quad \epsilon_0=-\frac{J}{4}(2+\Delta)\nonumber \\
&|\psi_1(j)\rangle=\frac{1}{\sqrt{2}}(|0_j 1_{j+1} \rangle+ | 1_j 0_{j+1}\rangle),\quad \epsilon_1=\frac{J}{4}(2-\Delta) \nonumber \\
&|\psi_2(j)\rangle=\frac{1}{\sqrt{2}}(|0_j 0_{j+1} \rangle + |1_j 1_{j+1}\rangle),\quad \epsilon_2=\frac{J}{4}\Delta \nonumber \\
&|\psi_3(j)\rangle=\frac{1}{\sqrt{2}}(|0_j 0_{j+1} \rangle- |1_j 1_{j+1}\rangle),\quad \epsilon_3=\frac{J}{4}\Delta
\label{eq:Bell}
\end{align}
where the corresponding energy eigenvalues are denoted by $\epsilon_\mu,\,\mu=0,1,2,3$. 
We use $\ket{0}$ to denote the spin-up state, and $\ket{1}$ the spin-down state.

The initial state can then be written as a product state:
\be
%|\Psi_0 \rangle=\prod_{i=1}^n |\psi_0(2i-1)\rangle\;
|\Psi_0 \rangle=\bigotimes_{j=1}^n |\psi_0(2j-1)\rangle\;
\ee
Similarly, the $k$-th eigenstate $\Phi_k$ ($k=1,2,\dots,2^{2n}$) of the postquench Hamiltonian ${\cal H}(0,J,\Delta)$ is given by:
\be
%|\Phi_k \rangle=\prod_{i=1}^n |\psi_{\mu_k}(2i)\rangle,\quad \mu_k=0,1,2,3\;
%|\Phi_k \rangle=\prod_{i=1}^n |\psi_{i_k}(2i)\rangle,\quad i_k=0,1,2,3\;
|\Phi_k \rangle=\bigotimes_{j=1}^n |\psi_{j_k}(2j)\rangle,\quad j_k=0,1,2,3\;
\label{eq:Phi_k}
\ee
with corresponding energies
\be
      E_k = \sum_{j=1}^n \epsilon_{j_k},\quad j_k = 0,1,2,3\,.
      \label{eq:E_k}
\ee
Expanding $|\Psi_0 \rangle$ in terms of $|\Phi_k \rangle$:
\be
|\Psi_0 \rangle=\sum_k a_k |\Phi_k \rangle\;,
\label{eq:expansion}
\ee
we obtain the time-evolved state
\begin{align}
    \ket{\Psi_0(t)}& = \sum_{k=1}^{\mathcal{N}} a_k(t) \ket{\Phi_k} \nonumber\\ 
		   & = \sum_{k=1}^{\mathcal{N}} a_k \exp(-i E_k t) \ket{\Phi_k}\,,
    \label{eq:Psi_t}
\end{align}
where $\mathcal{N}=2^{2n}$. The expansion coefficients $a_k=\skp{\Phi_k}{\Psi_0}$, 
given by scalar products between products of Bell states and singlet states, are all real. 
Once the set of $\Delta$-independent coefficients $a_k$ is known, the time-dependent state at any $t$ 
under the postquench Hamiltonian with arbitrary anisotropy parameter $\Delta$ can be determined.

\subsection{Expansion coefficients $a_k$}
\label{sec:a_k}

The scalar product for a coefficient $a_k$ can be written in terms of spin states in $Z$-basis 
$\ket{Z}=\ket{z_1 z_2 \cdots z_N}$ as
\be
      %a_k=\skp{\Phi_k}{\Psi_0}=\frac{1}{2^n} \sum_{\alpha=1}^{n}\skp{z'_1 z'_2 \cdots z'_N}{z_1 z_2 \cdots z_N}_\alpha \sigma
      a_k=\skp{\Phi_k}{\Psi_0}=\frac{1}{2^n} \sum_{Z' Z} \skp{Z'}{Z} \sigma'\sigma\,,
      \label{eq:ak_spins}
\ee
where $\sigma$ and $\sigma'$ are the signs of the states $\ket{Z}$ and $\ket{Z'}$, respectively.
In this expression, only the terms with $Z=Z'$ contribute.

%%%%%%%%%%% FIG1 %%%%%%%%%%%

\begin{figure}[t]
\includegraphics[width=8.6cm]{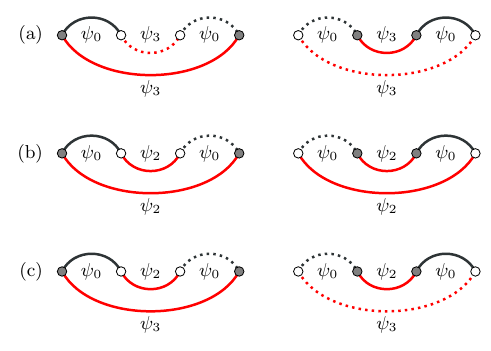}  % spin_loop.pdf
\caption{\label{fig:loop} Overlap diagrams for a positive (a), negative (b) and zero (c) coefficient $a_k=\skp{\Phi_k}{\Psi_0}$ in a four-site chain with PBC. 
Gray and white circles represent spin-up state $\ket{0}$ and spin-down state $\ket{1}$, respectively.
Black and red arches represent valence bonds (Bell dimers) $\psi_i$ 
in the states $\ket{\Psi_0}$ and $\ket{\Phi_k}$, with solid (dotted) arches
indicating positive (negative) contributions determined by the spin orientations. 
In each  panel, the right diagram is obtained from the left by a global spin inversion.
In (a), two negative dimers remain under spin inversion, resulting in a positive coefficient $a_k = 2\cdot 2^{-2}$; 
in (b), one negative dimer remains, corresponding to a negative coefficient $a_k = -2\cdot 2^{-2}$; 
in (c) the number of negative dimers change from one to two after spin inversion, representing a case with zero coefficient $a_k=0$.}
\end{figure}

%%%%%%%%%%%%%%%%%%%%%%%%%%

Alternatively, $a_k$ can be formulated as the overlap between the two states
$\ket{\Phi_k}$ and $\ket{\Psi_0}$ in the valence-bond basis. 
A valence bond here refers to a singlet or triplet dimer, i.e. a Bell state. 
In this paper, this basis is also referred to as the Bell basis. 
For a given dimer configuration in $\ket{\Phi_k}$ ($\ket{\Psi_0}$ is a fixed set of singlet dimers) in a chain with
PBC,   the overlap is determined by a single closed loop in the transition
graph in which the singlet-triplet dimers of two states  are
superimposed~\cite{Sutherland,Beach} [cf. Fig.~\ref{fig:loop}].  The spin configurations on the loop must
be consistent with the dimer configurations in both  $\ket{\Phi_k}$ and
$\ket{\Psi_0}$.  There exist two compatible spin configurations related by
global spin inversion, resulting in possible finite overlaps: $a_k = \pm 2\cdot 2^{-n}$, 
where the factor $2^{-n}$ arises from the normalization in Eq.~(\ref{eq:ak_spins}).

The sign of the overlap depends on the dimer types and the spin orientations
along the loop. For instance, a dimer in the triplet state $\ket{\psi_3}=(\ket{00}-\ket{11})/\sqrt{2}$ 
contributes positively when connecting two up-spins, but negatively when
connecting two down-spins. 
Consequently, a loop containing an even (odd) number of
such negative dimers produces a positive (negative) overlap. In certain cases, the
overlap vanishes because the loop changes sign under a global spin flip.
Figure~\ref{fig:loop} illustrates some cases for positive, negative and zero overlaps (coefficients).  

Since all nonzero coefficients are of equal magnitude $\abs{a_k}=2^{-(n-1)}$,
normalization of  $\ket{\Psi_0(t)}$ requires exactly $\mathcal{N}^*=4^{n-1}$ nonzero coefficients.
Thus, the coefficients $a_k$ are given by
\be
a_k=
\begin{cases}
\pm 2^{-(n-1)},& {\rm for}~ k \in k^*=1,2,\dots,4^{n-1} \\
0,& {\rm otherwise}\;,
\label{eq:a}
\end{cases}
\ee
where $k^*$ denotes the set of indices for nonzero $a_k$.

By encoding the Bell states as $(\psi_0, \psi_1, \psi_2, \psi_3)=(0,1,2,3)$,
we have identified two selection rules for all dimer configurations in $\ket{\Phi_k}$ [Eq.~(\ref{eq:expansion})]
with nonzero $a_k$:
\begin{itemize}
\item[(1)] the sum of all dimer-type indices $\sum_{j=1}^n \psi_{j_k}$ must be even; and 
%\item[(2)] the total number of $\psi_2$ and $\psi_3$ dimers must also be even.  
\item[(2)] the combined count of $\psi_2$ and $\psi_3$ dimers must also be even.
\end{itemize}

\subsection{Dynamical quantities}   

We are interested in the time evolution of the entanglement entropy and
the Loschmidt echo. The former characterizes how entanglement spreads through the system 
after a quantum quench; the latter serves as the observable for identifying dynamical phase transitions. 

The entanglement entropy (EE) quantifies the degree of quantum entanglement between a subsystem and its complement.
For a time-evolved state $\ket{\Psi_0(t)}$ partitioned into two parts $A$ and $B$, 
the entanglement between them is encoded
in the reduced density matrix of one part, say $A$,
\be
      \rho_A(t) = \text{Tr}_B \ket{\Psi_0(t)}\bra{\Psi_0(t)}\,,
\ee
where $\text{Tr}_B$ denotes the partial trace over $B$.
We consider both the von Neumann EE, defined by
\be
      \mathcal{S}_1(t) = -\text{Tr} \left[\rho_A(t) \log_2 \rho_A(t)  \right]\,,
      \label{eq:S1}
\ee
and the second-order R\'enyi EE, given by
\be
      \mathcal{S}_2(t) = -\log_2 \text{Tr} \left[ \rho_A^2(t) \right]\,.
      \label{eq:S2}
\ee 

For the quench protocols studied here, both the von Neumann and  R\'enyi EE
exhibit persistent oscillations, indicating the absence
of relaxation due to confinement~\cite{Calabrese_nphys,Viti}.
Such persistent entanglement oscillations have been previously reported in
the fully dimerized XX chain with $\Delta=0$ (equivalently, the flat-band 
Su-Schrieffer-Heeger chain)~\cite{Ueda,XX}. Here we extend this analysis to
the XXZ chain with $\Delta \neq 0$.

The Loschmidt echo $\mathcal{L}(t)$, also known as the return probability,  quantifies the overlap between the initial state $\ket{\Psi_0}$ and 
its time-evolved counterpart $\ket{\Psi_0(t)}$~\cite{Heyl,Karrasch,Heyl2014,Sirker,Vajna,DQPT_rev,Heyl_rev}: 
%\be
%       \mathcal{L}(t) = \abs{\skp{\Psi_0}{\Psi_0(t)}}^2 = \abs{\trip{\Psi_0}{\exp(-i\mathcal{H}_2(J,\Delta) t)}{\Psi_0}}^2\,. 
%       \label{eq:L}
%\ee
\be
      \mathcal{L}(t) = \abs{\mathcal{G} (t)}^2,\quad \mathcal{G} (t) = \skp{\Psi_0}{\Psi_0(t)}\,.
       \label{eq:L}
\ee
In the limit of large $N$, the Loschmidt echo has an asymptotic form given by
\be
      \mathcal{L}(t) = \exp(-N r(t))\,,
\ee
where $r(t)$ expressed as
\be
    r(t)=-\frac{1}{N} \ln  \mathcal{L}(t)\,,
    \label{eq:return}
\ee
is known as the return rate function.

The Loschmidt echo can be interpreted as the squared magnitude of an
out-of-equilibrium analog of the partition function in equilibrium statistical
mechanics. In the thermodynamic limit, zeros in the Loschmidt echo signal a
dynamical phase transition, with singularities appearing in the return rate
function~\cite{Heyl,Karrasch,Heyl2014,Sirker,Vajna,DQPT_rev,Heyl_rev}. Exact zeros of the Loschmidt echo can also occur in finite-size
systems under certain conditions, as reported in recent studies~\cite{Finite_zeros,Finite_zeros1,Finite_zeros2}. 
In this work, we identify finite-size Loschmidt zeros in our quench experiments for specific anisotropy parameters.

\section{Exact results}
\label{sec:exact}

Here we show the results for finite chains, obtained by evaluating the
expressions in Eqs.~(\ref{eq:S1}), (\ref{eq:S2}) and (\ref{eq:L}).
%We have obtained closed-form solutions for chains of lengths up to $N=2n=12$.
%For longer chains, exact results can also be obtained   
%by operating in the Bell basis. 

\subsection{Half-chain entanglement entropy}
\label{sec:exact_S}

We consider the entanglement between two halves of the chain with PBC,
quantified by the von Neumann entropy and the second-order R\'enyi entropy.
Without loss of generality, we set one boundary between site $i=n$ and $i=n+1$
and the other between site $i=2n$ and $i=1$. 
%For even $n$, the boundaries are
%then between unconnected sites of the initial state $\ket{\Psi_0}$, while for
%odd $n$ one  boundary is on a singlet link and the other is between two
%unconnected sites, as  illustrated in Fig.~\ref{fig:partition}.

The key ingredient of the entanglement entropy is 
the reduced density matrix, $\rho_A$, of the subsystem considered.
Here for a half chain, the dimension of the reduced density matrix is $2^n \times 2^n$.
Using the eigenvalues of $\rho_A$, that are all real and satify
$0 \le \lambda_l(t) \le 1,\;\sum_{l=1}^{2^n}\lambda_l(t)=1$,
we obtain the von Neumann EE by
\be
{\cal S}_1(t)=-\sum_{l=1}^{2^n} \lambda_l(t) \log_2[\lambda_l(t)]\,,
\label{eq:entropy_def1}
\ee
and the R\'enyi EE by
\be
{\cal S}_2(t)=-\log_2 \sum_{l=1}^{2^n} \lambda^2_l(t)\;.
\label{eq:entropy_def2}
\ee

\subsubsection{$n=2$}
We start with the shortest chain with $n=2$. 
There are four nonzero coefficients $a_1=a_2=a_4=1/2$ and $a_3=-1/2$ contributed by the 
following dimer configurations in $\ket{\Phi_k}$:
\begin{align}
       \ket{\Phi_1} &= \ket{\psi_0(2)}\ket{\psi_0(4)}\,, \nonumber \\
       \ket{\Phi_2} &= \ket{\psi_1(2)}\ket{\psi_1(4)}\,, \nonumber \\
       \ket{\Phi_3} &= \ket{\psi_2(2)}\ket{\psi_2(4)}\,, \nonumber \\
       \ket{\Phi_4} &= \ket{\psi_3(2)}\ket{\psi_3(4)}\,, 
\label{eq:dimers_n2}
\end{align}
with energies $E_1=-J(2+\Delta)/2,\,E_2=J(2-\Delta)$ and $E_3=E_4=J\Delta/2$.
The time-evolved state $\ket{\Psi_0(t)}$ in the spin basis
is then given by 
\begin{align}
|\Psi_0(t) \rangle= \frac{1}{4}&(1/e^+ + e^-)\left(|1 0 1 0 \rangle+
|0 1 0 1 \rangle\right) \nonumber \\
 -\frac{1}{2}&e^0\left(|1 0 0 1\rangle + |0 1 1 0 \rangle\right) \nonumber \\
 +\frac{1}{4}&(-1/e^+ + e^-)\left(|1 1 0 0\rangle + |0 0 1 1 \rangle \right)\;,
\label{eq:Psit_n2}
\end{align}
where the shorthand notations  $e^{\pm}=\exp\left[-i \left(1 \pm \Delta/2\right)Jt\right]$ and $e^{0}=\exp\left[-i (\Delta/2) Jt \right]$ are used. 
The reduced density matrix for the first half of the chain in terms of the basis
$\ket{00}, \ket{01}, \ket{10}$ and $\ket{11}$ is given by:
\be
  \begin{bmatrix}
\frac{1}{4}\sin^2 Jt & 0 & 0 & 0\\
0 & \frac{1} {4}(1+\cos^2 Jt)  & -\frac{1}{2}\cos Jt\cos \Delta Jt & 0\\
0  & -\frac{1}{2}\cos Jt\cos \Delta Jt & \frac{1}{4}(1+\cos^2 Jt)  & 0\\
0& 0 & 0 & \frac{1}{4}\sin^2 Jt \\
\end{bmatrix}\,,
\ee
which has the eigenvalues:
\begin{align}
\lambda_{1,2}&=\frac{1}{4}\sin^2(Jt)\nonumber \\
 \lambda_{3,4}&=\frac{1}{4}[1+\cos^2(Jt) \pm 2 \cos(Jt)\cos(J\Delta t)]\;.
 \label{eq:lambda_n2}
 \end{align}
Thus, the von Neumann EE can be obtained by inserting the eigenvalues into Eq.~\eqref{eq:entropy_def1},
yielding
\begin{align}
  {\cal S}_1^{n=2}(t) & = 2 - \frac{1}{2} \sin^2(Jt) \log_2 \sin^2(Jt) \nonumber \\
                      & -\frac{1}{4} \left[ 1+\cos^2(Jt) + 2\cos(Jt) \cos(J\Delta t)  \right] \nonumber \\
                      & \times \log_2 \left[ 1+\cos^2(Jt) + 2\cos(Jt) \cos(J\Delta t)  \right]   \nonumber \\
 		      &  -\frac{1}{4} \left[ 1+\cos^2(Jt) - 2\cos(Jt) \cos(J\Delta t)  \right] \nonumber \\	 
		      & \times \log_2 \left[ 1+\cos^2(Jt) - 2\cos(Jt) \cos(J\Delta t)  \right]\,.
  \label{eq:S1_n2}
\end{align}
Similarly, the  R\'enyi EE is obtained from Eq.~\eqref{eq:entropy_def2}:
\begin{align}
{\cal S}_{2}^{n=2}&(t)=3 - \log_2\left[ 1+ \sin^4(Jt)+2\cos^2(Jt)\right.\nonumber \\
&\left. +\cos^4(Jt)+4\cos^2(Jt)\cos^2(J \Delta t)\right]\;.
\label{eq:S2_n2}
\end{align}
Both functions ${\cal S}_1^{n=2}(t)$ and ${\cal S}_2^{n=2}(t)$ reach their maximum value of $2$, independent of $\Delta$, at 
$t=(2m+1)\pi/2J$ with $m=0,1,2,\cdots$ when the condition $\cos(Jt)=0$ is satisfied.

For the XX-chain with $\Delta=0$, the expression for the von Neumann EE is reduced to
%\be
\begin{align}
{\cal S}_{1,\text{XX}}^{n=2}(t) &=-2\left[\sin^2\bigl(\frac{Jt}{2}\bigr) \log_2 \sin^2\bigl(\frac{Jt}{2}\bigr) \right. \nonumber \\
                                &+ \left. \cos^2\bigl(\frac{Jt}{2}\bigr) \log_2 \cos^2 \bigl(\frac{Jt}{2}\bigr)  \right]\,,
\label{eq:S1XX_n2}
\end{align}   
%\ee
and for the R\'enyi EE is
\be
  {\cal S}_{2,\text{XX}}^{n=2}(t) = -2\log_2 \left[1- \sin^2(Jt)/2 \right]\,,
\ee
as also known from free-fermion calculations~\cite{Eisler,XX}.

\subsubsection{$n=3, 5, \cdots$}

%%%%%%%%%%% FIG2 %%%%%%%%%%%

\begin{figure}[t]
\includegraphics[width=8.6cm]{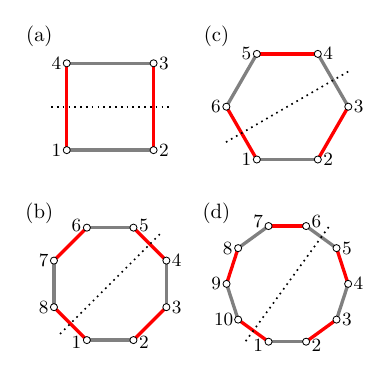} % bipartition.pdf
\caption{\label{fig:partition} Sketch of the bipartition for chains with PBC and $n=2, 3, 4, 5$. Gray lines denote the singlet dimers in the initial
state $\ket{\Psi_0}$, and red lines represent the couplings of the postquench Hamiltonian. 
The dotted line in each figure marks a symmetric bipartition of the chain.
(a) and (b): For even $n$, the bipartition cuts two couplings of the post-quench Hamiltonian. 
(c) and (d): For odd $n$, the bipartition cuts one postquench coupling and one dimer in the initial state.}
\end{figure}

%%%%%%%%%%%%%%%%%%%%%%%%%%

For $n=3$, the eigenvalues of the $8\times 8$ reduced density matrix are
\begin{align}
&\lambda_{1,2,3,4}=\frac{1}{8}\sin^2\bigl(\frac{Jt}{2} \bigr) \nonumber\\
&\lambda_{5,6}=\frac{1}{8}\left[1+\cos^2 \bigl(\frac{Jt}{2} \bigr) + 2 \cos{\bigl(\frac{Jt}{2} \bigr)} \cos\bigl( \frac{J\Delta t}{2} \bigr) \right]\nonumber\\
&\lambda_{7,8}=\frac{1}{8}\left[1+\cos^2\bigl( \frac{Jt}{2} \bigr)-  2 \cos{\bigl(\frac{Jt}{2} \bigr)} \cos\bigl( \frac{J\Delta t}{2} \bigr)\right]\;.
\label{eq:lambda_n3}
\end{align}
These eigenvalues at $t$ are related to those of the $n=2$ case [see Eq.~\eqref{eq:lambda_n2}] by
\be
      \lambda^{n=3}(t)  = \frac{1}{2} \lambda^{n=2}(t/2)\,.
\ee 
Consequently, the entanglement entropies $\mathcal{S}_1$ and $\mathcal{S}_2$  for $n=2$ and $n=3$ 
satisfy
$
{\cal S}^{n=3}(t)=1 +{\cal S}^{n=2}(t/2)\;.
\label{eq:S23}     
$
More generally, this relation holds for any odd $n$,
\be
{\cal S}^{n=\text{odd}}(t)=1 +{\cal S}^{n=2}(t/2)\;,
\label{eq:S_odd}
\ee
and arises from the behavior of the local states at the boundaries of the bipartition.

To clarify this,  Fig.~\ref{fig:partition} illustrates the initial singlet-product state (gray lines) and the couplings (red lines) 
in the postquench Hamiltonian for chains of length $n=2,4$ (Fig.~\ref{fig:partition}(a) and (b)) and $n=3,5$  (Fig.~\ref{fig:partition} (c) and (d)), where the dashed line indicates the symmetric bipartition.
In our setup, each initial entangled dimer located at $(i, i+1)$ 
can spread over three lattice spacings, covering sites from $i-1$ to $i+2$, during the time evolution~\cite{Eisler,XX}. 
The EE of a half chain is the number of entangled states that connect sites inside the partition to sites outside.
For any odd $n>2$, the initial state always contains an entangled dimer that crosses the boundary between sites $i=n$ and $i=n+1$, 
contributing a time-independent unit to the EE;
this is the constant term '$1$' in Eq.~\eqref{eq:S_odd}.
In addition, at most two entangled states may cross the boundary between sites $i=1$ and $i=2n$,
evolved from the singlets initially at $(1,2)$ and $(2n-1,2n)$.
This time-dependent part of the EE is analogous to the $n=2$ case. 
However, in the $n=2$ chain, the short chain length forces a single entangled state 
to cross both boundaries simultaneously, whereas for $n>2$ it can cross only one boundary.
As a result, one unit of the EE in the $n=2$ case 
involves two couplings with a total strength $2J$ in the postquench Hamiltonian (red lines in Fig.~\ref{fig:partition}),
while for $n>2$ it involves only a single postquench coupling of strength $J$. 
This difference leads to a doubling of the entanglement oscillation period for odd $n>2$
compared with the $n=2$ case.

\subsubsection{$n=4,6,\cdots$}

%%%%%%%%%%% FIG3 %%%%%%%%%%%

\begin{figure}[t]
\includegraphics[width=8.6cm]{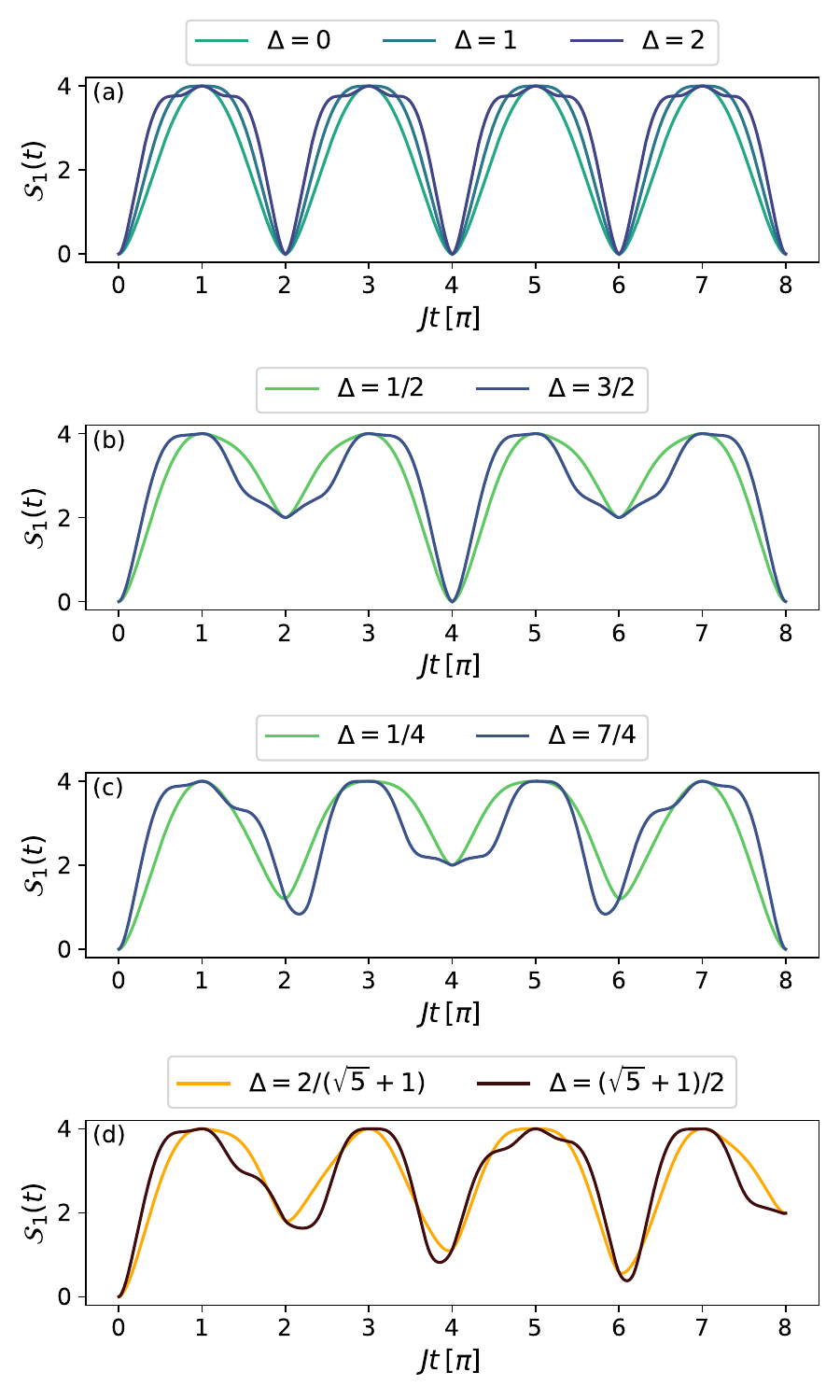} % exact_S_even_h.pdf
\caption{\label{fig:S_even} Time dependence of the half-chain von Neumann entanglement entropy for the XXZ chain with even $n\ge 4$ at 
various values of the anisotropy parameter $\Delta$. All curves are plots of the analytical results using Eq.~\eqref{eq:S_even}.}
\end{figure}

%%%%%%%%%%%%%%%%%%%%%%%%%%

Repeating the above reasoning, we can also find an exact relation between the dynamical EE of even $n>2$ and that of $n=2$.
Similar to the $n=2$ case, no singlet-dimer initially crosses the bipartition boundaries.
The dynamical EE arises solely from two pairs of dimers initially located adjacent to the 
boundaries, one pair at $(n,n-1)$ and $(n+1,n+2)$, and the other at $(1,2)$ and $(2n-1, 2n)$.
For $n>2$, each pair of dimers is connected through one postquench coupling of strength $J$, and each dimer can cross only one
boundary during the time evolution. 
In contrast, in a $n=2$ chain there exists only one such pair coupled through
two postquench couplings ($2J$), and each entangled state crosses the two boundaries at the same time.  
Therefore, the dynamical EE for even $n>2$ is related to that of the $n=2$ case by
\be
{\cal S}^{n=\text{even}}(t)=2{\cal S}^{n=2}(t/2)\;,
\label{eq:S_even}       
\ee
where the prefactor 2 reflects the doubled number of contributing dimers for $n>2$,
and the factor $1/2$ in the time dependence reflects the fact that
only one postquench coupling is involved in each pair of dimers.

From Eq.~\eqref{eq:S_even}, the von Neumann EE for an even $n>2$ chain is explicitly given by
\begin{align}
  {\cal S}_1^{n=\text{even}}&(t)  = 4 - \sin^2(Jt/2) \log_2 \sin^2(Jt/2) \nonumber \\
                      & -\frac{1}{2} \left[ 1+\cos^2(Jt/2) + 2\cos(Jt/2) \cos(J\Delta t/2)  \right] \nonumber \\
                      & \times \log_2 \left[ 1+\cos^2(Jt/2) + 2\cos(Jt/2) \cos(J\Delta t/2)  \right]   \nonumber \\
                      &  -\frac{1}{2} \left[ 1+\cos^2(Jt/2) - 2\cos(Jt/2) \cos(J\Delta t/2)  \right] \nonumber \\
                      & \times \log_2 \left[ 1+\cos^2(Jt/2) - 2\cos(Jt/2) \cos(J\Delta t/2)  \right]\,,
  \label{eq:S1_even}
\end{align}
and the R\'enyi EE is
\begin{align}
{\cal S}_{2}^{n=\text{even}}&(t)=6 - 2\log_2\left[ 1+ \sin^4(Jt/2)+2\cos^2(Jt/2)\right.\nonumber \\
&\left. +\cos^4(Jt/2)+4\cos^2(Jt/2)\cos^2(J \Delta t/2)\right]\;.
\label{eq:S2_even}
\end{align}
For $n=4$, the analytical forms of the eigenvalues of the $16\time 16$ reduced density matrices for the XX- and XXX-chains 
are provided in Appendix~\ref{sec:S_n4}.

We note that the dynamical entanglement entropy is a periodic function of time, with a period $2q\pi$ for $n>2$ and $q\pi$ for $n=2$,
provided $J\Delta = p/q$ with $p$ and $q$ being coprime integers.
To illustrate this, we plot in Fig.~\ref{fig:S_even} the time-dependent entanglement entropy for chains with even $n>2$
and with various values of the anisotropy parameter $\Delta$ by fixing $J=1$. 
The case for the golden mean values %$\Delta =\tau$ and $\Delta=1/\tau$ with $\tau=(\sqrt{5}+1)/2$,
$\Delta = 2/(\sqrt{5}+1)$ and $\Delta=(\sqrt{5}+1)/2$ 
shown in Fig.~\ref{fig:S_even}(d), displays no periodicity.

\subsection{Loschmidt echo and return rate function}
\label{sec:echo}

Given the $2n$-site initial state and the time-evolved state in the Bell representation [Eq.~\eqref{eq:Psi_t}],
the Loschmidt echo, defined as the overlap of both states, can be rewritten as
\begin{align}
     \mathcal{L}(t) &=\left| \sum_{k\in k^*}  a_k^2 \exp(-iE_k t)\right|^2 \nonumber \\
                    &=\left|\frac{1}{4^{n-1}}  \sum_{k\in k^*} \exp(-iE_k t) \right|^2\,,
     \label{eq:exact_echo}
\end{align}
where the nonzero coefficients $a_{k}$ satisfy the selection rules given in Sec.~\ref{sec:a_k},
and the energy $E_{k}$ of $k$-th state is just the sum of  the energies
of participating Bell dimers [Eq.~\eqref{eq:E_k}].

%%%%%%%%%%% FIG4 %%%%%%%%%%%

\begin{figure}[t]
\includegraphics[width=8.6cm]{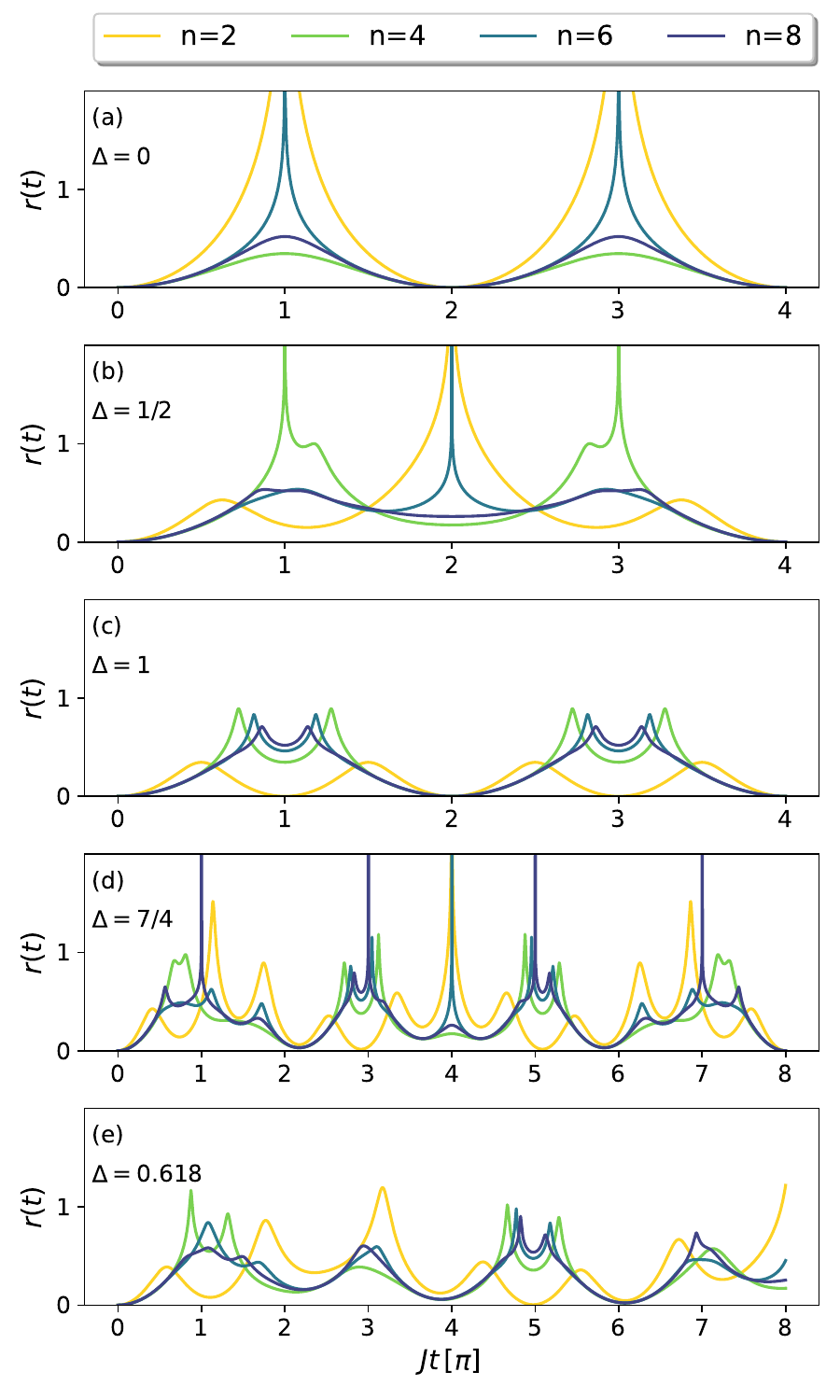} %exact_return_even_h.pdf
\caption{\label{fig:return_even} Time dependence of the return rate function for the XXZ chain of length $N=2n,\,n=2, 4, 6, 8$ and with the anisotropy parameter 
at (a) $\Delta = 0$, corresponding to the XX chain; (b) $\Delta =1/2$; (c) $\Delta=1$, the XXX chain; (d) $\Delta=7/4$; 
(e) an irrational $\Delta=2/(\sqrt{5}+1)\approx 0.618$.}
\end{figure}

%%%%%%%%%%%%%%%%%%%%%%%%%%

%%%%%%%%%%% FIG5 %%%%%%%%%%%

\begin{figure}[t]
\includegraphics[width=8.6cm]{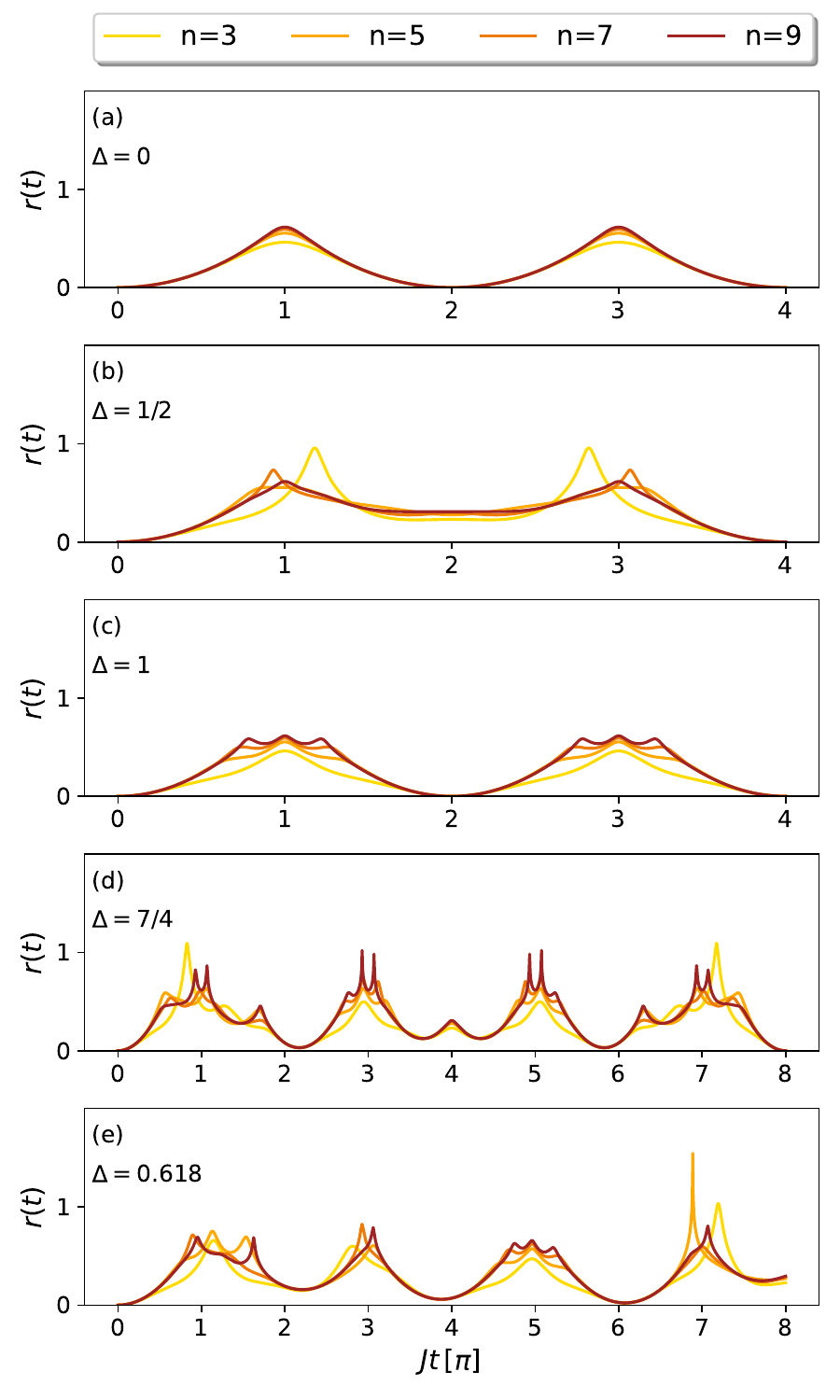}  % exact_return_odd_h.pdf
\caption{\label{fig:return_odd} Time dependence of the return rate function for the XXZ chain with odd $n=3, 5, 7, 9$ and
with the anisotropy parameter at (a) $\Delta = 0$, corresponding to the XX chain; 
(b) $\Delta =1/2$; (c) $\Delta=1$, the XXX chain; (d) $\Delta=7/4$; (e) an irrational $\Delta=2/(\sqrt{5}+1)\approx 0.618$.}
\end{figure}

%%%%%%%%%%%%%%%%%%%%%%%%%%

%Below we present closed-form solutions of short XXZ chains with $n=2, 3, 4, 6$ and XX chains up to $n=8$.
For the shortest chain with $n=2$, the dimer configurations for  nonzero coefficients are given in
Eq.~\eqref{eq:dimers_n2}, leading to the Loschmidt echo
\begin{align}
{\cal L}^{n=2}(t)&=\frac{1}{16}\left|e^+ +2 e^0 + e^-\right|^2\cr
&=\frac{1}{4}\left [1+ 2\cos(Jt) \cos(\Delta Jt) + \cos^2(Jt) \right]\;.
\label{eq:L_n2}
\end{align}
where the same shorthand notations as in Eq.~\eqref{eq:Psit_n2} are used.

For the $n=3$ chain with 16 nonzero $a_k$,  the analytical result for Loschmidt echo is
\begin{align}
{\cal L}^{n=3}(t)&=\frac{1}{256}\left\{36\cos(\Delta Jt)+36\cos[(1-\Delta)Jt] + \right. \nonumber\\
&\left. +12\cos[(1+\Delta)Jt]+72 \cos(Jt)  \right. \nonumber\\
&\left. +6 \cos(2Jt)+12\cos[(2+\Delta)Jt] + 82\right\}\;.
\label{eq:L_n3}
\end{align}
For the chain with $n=4$, we obtain
\begin{align}
{\cal L}^{n=4}(t)=\frac{1}{256}&\left\{\left[12 \cos^2(Jt/2) +  \cos(J\Delta t)\left(3+\cos^2 (Jt)\right)\right]^2\right. \nonumber\\
&\left. +\sin^2(J\Delta t)\sin^4 (Jt) \right\}\;.
\end{align}
%
%and for $n=6$ with an arbitrary $\Delta$ the Loschmidt echo is given in Appendix~\ref{sec:echo_n6}.
%An analytical expression for $n=6$ with arbitrary values of $\Delta$ is given in Appendix~\ref{sec:echo_n6}. 

%For the special case $\Delta=0$ (XX chain), we have
%%
%\begin{align}
%{\cal L}_{XX}^{n=2}(t)=\frac{1}{4}&\left[1+\cos (Jt) \right]^2  \;,\\
%{\cal L}_{XX}^{n=4}(t)=\frac{1}{4^{5}}&\left[19+12\cos (Jt)+\cos(2Jt) \right]^2 \;, \\
%%{\cal L}_{XX}^{n=6}(t)=\frac{1}{4^8}&\left[ 113 + 128\cos(Jt) + 15\cos(2Jt) \right]^2\;,\\
%{\cal L}_{XX}^{n=6}(t)=\frac{1}{4^{9}}&\left[226+255\cos (Jt))+30\cos(2Jt)\right. \nonumber\\
%&\left. +\cos(3Jt) \right]^2 \;,\\
%{\cal L}_{XX}^{n=8}(t)=\frac{1}{4^{13}}&\left[3225+3976\cos (Jt)+924\cos(2Jt)\right. \nonumber\\
%&\left. +56\cos(3Jt)+\cos(4Jt) \right]^2 \;,
%\end{align}
%%
%where the Loschmidt echo for even $n$ is written as a quadratic form in a superposition of $\cos(\mu Jt)$-terms
%with $\mu=0,1,\cdots,n/2$ and a global prefactor $C_n=1/4^{2n-3}$. 
%Similarly, for odd $n$, it can be written as linear combination of $\cos(\mu Jt)$-terms with 
%$\mu=0,1,\cdots,n-1$ and $C_n=1/4^{2n-2}$.  
%Explicit expressions for odd $n=5,7$ and $9$ are provided in Appendix~\ref{sec:echo_n579}.
%Extensions to larger $n$ are straightforward.
%These expressions also imply that the corresponding Loschmidt amplitude, $\mathcal{G}(t)=\skp{\Psi_0}{\Psi_0(t)}$, is real for even $n$
%and is complex for odd $n$.
%%

In Figs.~\ref{fig:return_even} and \ref{fig:return_odd} we plot the rate function $r(t)$ 
(defined in Eq.~\eqref{eq:return}) of the Loschmidt echo as functions
of time for various chain lengths and five values of the anisotropy parameter, with fixed $J=1$.
Consistent with the dynamical entanglement entropy, the rate function for $n>2$ is a periodic function of time 
with a period of $2 q \pi$, provided $J\Delta = p/q$ with $p$ and $q$ being coprime integers. 
For the shortest chain, $n=2$, the period deviates from this rule in some cases, such as when $\Delta=1$,
where the period is $\pi$  [Fig.~\ref{fig:return_even}(c)].
For an irrational value of $\Delta$,
the rate function is non-periodic, as illustrated in Figs.~\ref{fig:return_even}(e) and \ref{fig:return_odd}(e)  
for $\Delta = 2/(\sqrt{5}+1)$.

Moreover, we observe exact zeros of the Loschmidt echo for even $n$, signaled by diverging peaks in the rate function. For instance,
the XX chain ($\Delta = 0$)  with $n=2$ has Loschmidt zeros at  $t=\pi, 3\pi, \cdots$; 
the $n=4$ XXZ chain with $\Delta=1/2$, and also the longer $n=8$ chain with $\Delta=7/4$
exhibit zeros at $t=\pi, 3\pi, \cdots$, too. 
At these instances, the return rate function becomes divergent, as shown in Fig.~\ref{fig:return_even}.
In all cases considered here, Loschmidt zeros do not appear in odd $n$ chains [Fig.~\ref{fig:return_odd}], 
nor in the XXX chain for any $n$. 
For the irrational value $\Delta=2/(\sqrt{5}+1)$, we find no Loschmidt zeros up to $t=8\pi$ 
[see Figs.~\ref{fig:return_even}(e) and \ref{fig:return_odd}(e)].

%\subsection{Dynamical quantum phase transitions and finite-size scaling}
\subsection{Dynamical quantum phase transitions}
\label{sec:dqpt}

%%%%%%%%%%% FIG6 %%%%%%%%%%%
%
%\begin{figure}[t]
%\includegraphics[width=8.3cm]{fig6.pdf}  % echo3_D_h.pdf % exact_return_infty_h.pdf
%\caption{\label{fig:r_infty} Return rate functions in the thermodynamic limit for five values of $\Delta$, where the label $\Delta=0.618$ represents an irrational value
%$\Delta=2/(\sqrt{5}+1)$.
%The solid black line in each subfigure is a plot of the analytical result in Eq.~\eqref{eq:r_infty},
%overlaid with the exact solutions (thin solid lines) for finite chains with $n=6, 8, 9$ dimers. Gray/red vertical dotted lines
%mark the $\Delta$-independent/dependent critical times. The horizontal dashed line indicates the maximum $r=\ln(2)$ for $n\to\infty$.
%Note that the rate function for $\Delta=1$ in the thermodynamic limit is a smooth function.}
%\end{figure}

Since the prequench and postquench Hamiltonians belong to distinct topological
phases, the quench considered here is expected to exhibit dynamical quantum phase
transitions (DQPTs)~\cite{Dora,Heyl_rev}. A DQPT is characterized by an exact zero of the Loschmidt echo
at the critical time or a nonanalytic cusp of the return rate function in the thermodynamic limit. 
As shown in Figs.~\ref{fig:return_even} and \ref{fig:return_odd},  
the rate functions in the finite systems, particularly with finite anisotropy $\Delta \neq 0$,  
display complex and highly nontrivial structures with multiple kinks whose behavior
appears to vary nonmonotonically with system size.
In this subsection we aim to understand 
how these kink structures  evolve as the system size increases and
whether they persist in the thermodynamic limit.

We first consider the XX case.
Under a fermionic transformation, the XX chain decouples into two
transverse-field Ising chains~\cite{XX-Ising1,XX-Ising2}. Within this mapping, the present protocol %by assuming $J=1$
then corresponds to a quench from a fully polarized initial state along the
transverse-field direction, $\ket{\Psi_0}=\bigotimes_i \ket{\rightarrow_i}$,
evolved under a classical Ising Hamiltonian. Equivalently, it can be viewed as
a quench from a fully polarized state into a cluster phase~\cite{Cluster} governed by 
${\cal H}_\text{cluster}=-J/4 \sum_i Z_{i-1} X_i Z_{i+1}$, where $X$ and $Z$ denote Pauli
matrices. Previous studies~\cite{Heyl115} have established that such quenches exhibit
DQPTs at critical times $t^*=(2m+1)\pi/J$, with $m=0,1,2,\cdots$.

%%%%%%%%%%%% FIG7 %%%%%%%%%%%%%%
%\begin{figure}[t]
%\includegraphics[width=8.3cm]{fig7.pdf}  % fisher_zeros.pdf
%\caption{\label{fig:fisher}
%Fisher zeros obtained by the condition $\abs{\lambda_+} = \abs{\lambda_-}$ in the complex time plane. (a) Isotropic case with $\Delta=1$: the curves only touches the real
%time axis (horizontal dashed line) tangentially; (b) $\Delta=7/4$: a DQPT occurs when the Fisher-zero line crosses the real axis. Gray/red vertical dotted lines
%mark the $\Delta$-independent/dependent critical times.}
%\end{figure}

To incorporate the effects of finite anisotropy $\Delta$, we derive below an analytical expression for the rate function in the limit $n\to \infty$.
We begin by obtaining a general expression for the Loschmidt amplitude $\mathcal{G}(t)=\trip{\Psi_0}{U(t)}{\Psi_0}$,
starting from the initial $n$ singlet pairings $\ket{\Psi_0} = \frac{1}{2^n} \bigotimes_j (\ket{0_{2j-1}1_{2j}}- \ket{1_{2j-1}0_{2j}})$.
Since the postquench Hamiltonian is  a sum of decoupled bond Hamiltonians with a wrap-around bond connecting the last and first sites, 
the time-evolution operator factorizes exactly as
$U(t) = \bigotimes_j U_{2j}(t)$ with $U_{2j}(t)=\exp(-it\,h_{2j})$.
The time-evolution operator $U_{2j}(t)$ acting on a single dimer $(2j, 2j+1)$ is given by:
\begin{align}
     U_{2j}(t) \ket{0 1} &=\quad e^{i\delta}\cos\theta \ket{0 1} - ie^{i\delta}\sin\theta \ket{1 0}\,,\label{eq:01} \\
     U_{2j}(t) \ket{1 0} &= - ie^{i\delta}\sin\theta \ket{0 1} + e^{i\delta}\cos\theta  \ket{1 0}\,, \label{eq:10}\\
     U_{2j}(t) \ket{0 0} &= e^{-i\delta} \ket{0 0}\,,  \label{eq:00} \\
     U_{2j}(t) \ket{1 1} &= e^{-i\delta} \ket{1 1}\,,   \label{eq:11}
\end{align}
with  $\theta = {Jt}/{2},\; \delta= {J\Delta t}/{4}$.
%\begin{align}
%     U_{2j}(t) \ket{0_{2j} 1_{2j+1}} &= C \ket{0_{2j} 1_{2j+1}} + S \ket{1_{2j} 0_{2j+1}}\,,\label{eq:01} \\
%     U_{2j}(t) \ket{1_{2j} 0_{2j+1}} &= S \ket{0_{2j} 1_{2j+1}} + C \ket{1_{2j} 0_{2j+1}}\,, \label{eq:10} \\
%     U_{2j}(t) \ket{0_{2j} 0_{2j+1}} &= D \ket{0_{2j} 0_{2j+1}}\,,  \label{eq:00} \\
%     U_{2j}(t) \ket{1_{2j} 1_{2j+1}} &= D \ket{1_{2j} 1_{2j+1}}\,,   \label{eq:11} 
%\end{align}
%where we use the short-hand notations: $C=\cos(Jt/2)e^{iJ\Delta t/4}$, $S=-i \sin(Jt/2)e^{iJ\Delta t/4}$,
%and $D=e^{-iJ\Delta t/4}$.
These identities imply that the dimers $\ket{0_{2j} 1_{2j+1}}$ and $\ket{1_{2j} 0_{2j+1}}$
undergo an oscillating spin-flip process characterized by the parameter $S\equiv - ie^{i\delta}\sin\theta$, whereas the
dimers $\ket{0_{2j} 0_{2j+1}}$ and $\ket{1_{2j} 1_{2j+1}}$ remain stationary.
Building on the Bell-state overlap analysis presented in the previous sections, we deduce an analytical 
expression for the Loschmidt amplitude of an $n$-dimer chain, 
denoting $C\equiv e^{i\delta}\cos\theta$
and $D \equiv e^{-i\delta}$:
\begin{align}
    \mathcal{G}^{(n)}(t) & = \frac{2}{2^{n}} \left [ \sum_{k=0}^{[n/2]} \binom{n}{2k} C^{n-2k} D^{2k} +S^n \right] \nonumber\\
                   & = \frac{1}{2^{n}} \left[ (C+D)^n + (C-D)^n + 2S^n \right]\,,
    \label{eq:G_dimer}
\end{align}
where $[n/2]$ is the integer part of $n/2$, and $\binom{n}{2k}$ are binomial coefficients.
In Eq.~\eqref{eq:G_dimer}, the factors $C^{n-2k}$ and $D^{2k}$ correspond to contributions from dimers on positions $(2j, 2j+1)$ 
with unequal and equal $z$-spin configurations, respectively, with the exponents representing their respective counts.
The term $S^n$  arises from the overlap between the N\'eel-like configurations $\ket{0101\cdots 01}$ and $\ket{1010\cdots 10}$,
which requires $n$ spin-flip processes.

The closed form of $\mathcal{G}^{(n)}(t)$ presented in Eq.~\eqref{eq:G_dimer} 
can also be obtained using the standard transfer matrix approach~\cite{Sirker,Heyl115}.
In this approach, the Loschmidt amplitude for a chain of $n$ dimers with PBC is given by   
$
     %\mathcal{G}(t) = \trip{\Psi_0}{U(t)}{\Psi_0} = \text{Tr}[T^n]\,,
      \mathcal{G}^{(n)}(t) = \text{Tr}[T^n]\,,
     %\label{eq:G_t}
$ 
in terms of a $4\times 4$ symmetric transfer matrix $T$ given by
\be
        T = \frac{1}{2}  \begin{bmatrix}
           e^{i\delta}\cos\theta  & 0                           & 0    & e^{-i\delta} \\
              0                        & ie^{i\delta}\sin\theta & 0    & 0 \\
              0                        &    0                        & ie^{i\delta}\sin\theta       & 0 \\
            e^{-i\delta}           & 0                           &   0  & e^{i\delta}\cos\theta  \\
        \end{bmatrix}\,,
     \label{eq:T_matrix}
\ee

%%%%%%%%%%% FIG6 %%%%%%%%%%%

\begin{figure}[t]
\includegraphics[width=8.3cm]{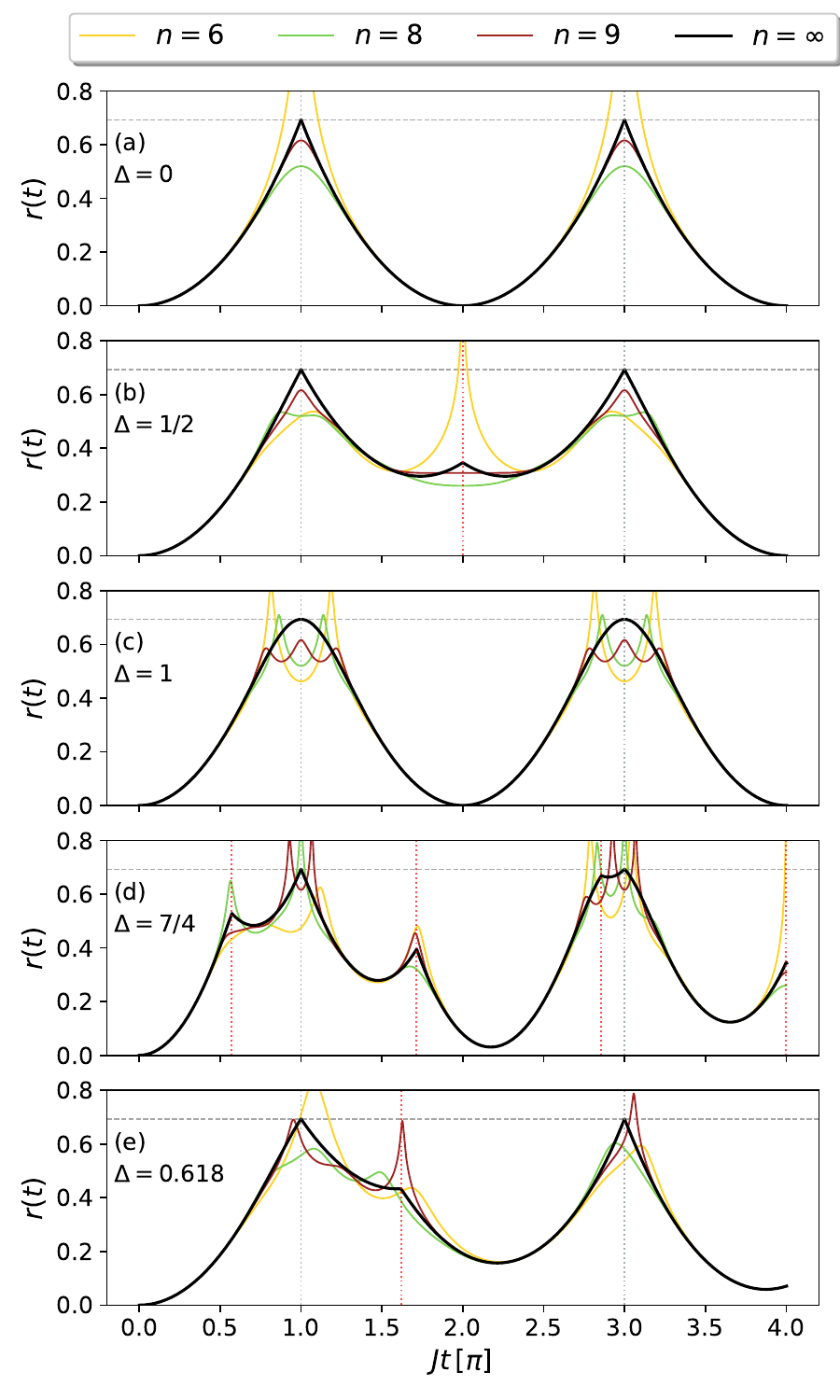}  % echo3_D_h.pdf % exact_return_infty_h.pdf
\caption{\label{fig:r_infty} Return rate functions in the thermodynamic limit for five values of $\Delta$, where the label $\Delta=0.618$ represents an irrational value
$\Delta=2/(\sqrt{5}+1)$.
The solid black line in each subfigure is a plot of the analytical result in Eq.~\eqref{eq:r_infty},
overlaid with the exact solutions (thin solid lines) for finite chains with $n=6, 8, 9$ dimers. Gray/red vertical dotted lines
mark the $\Delta$-independent/dependent critical times. The horizontal dashed line indicates the maximum $r=\ln(2)$ for $n\to\infty$.
Note that the rate function for $\Delta=1$ in the thermodynamic limit is a smooth function.}
\end{figure}
%%%%%%%%%%%%%%%%%%%%%

%with $\theta = {Jt}/{2},\; \delta= {J\Delta t}/{4}$.
The exact transfer matrix has four eigenvalues with a double degeneracy:
%\be
\begin{align}
      \lambda_1 & = \frac{1}{2} \left [e^{i\delta} \cos\theta  + e^{-i\delta}  \right] \equiv \lambda_+ \,, \\
      \lambda_2 & = \frac{1}{2} \left [e^{i\delta} \cos\theta  - e^{-i\delta}  \right] \equiv \lambda_- \,, \\
      %\lambda_1 & = \frac{1}{2}\cos\theta e^{i\delta} + \frac{1}{2}e^{-i\delta} \equiv \lambda_+ \,, \\
      %\lambda_1 & = \frac{1}{2} e^{-i\delta} \left [e^{2i\delta} \cos \theta  + 1 \right] \equiv \lambda_+ \,, \\
      %\lambda_2 & = \frac{1}{2} e^{-i\delta} \left [e^{2i\delta} \cos \theta   - 1 \right] \equiv \lambda_- \,, \\ 
      \lambda_3 & = \lambda_4 = \frac{i}{2} e^{i\delta} \sin\theta   \equiv \lambda_0 \,.
\end{align} 
For a finite chain with $n$ dimers, 
we reproduce the Loschmidt amplitude of the form:
\be
       \mathcal{G}^{(n)}(t) =  \lambda_+^n + \lambda_-^n + 2\lambda_0^n\,,
       \label{eq:G_finite}
\ee
identical to Eq.~\eqref{eq:G_dimer}.
In limit $n\to\infty$, the Loschmidt echo is then governed by the eigenvalue of largest magnitude provided there is no cancellation among $\lambda_+^n$,
$\lambda_-^n$ and $2\lambda_0^n$, leading to
\be
     \mathcal{L}^{\infty} = \abs{\mathcal{G}^\infty }^2 = \lim_{n\to \infty} \abs{\lambda_\text{max}}^{2n}\,.
     %\lambda_\text{max} = \begin{cases} \lambda_+ & \text{if }  \end{cases}
\ee
where
\begin{align}
     \abs{\lambda_\text{max}}^2 & = \max(\abs{\lambda_+}^2, \abs{\lambda_-}^2) \nonumber \\
                                & = \frac{1 + \cos^2 \theta + 2 \abs{\cos \theta \cos (2\delta)}}{4}\,. 
\end{align}
The degenerate eigenvalue $\lambda_0$ (corresponding to the spin-flip parameter $S$ in Eq.~\eqref{eq:G_dimer}) is never the dominant eigenvalue 
since
\begin{align}
      \max(\abs{\lambda_+}^2, \abs{\lambda_-}^2) - \abs{\lambda_0}^2 &= \frac{2\cos^2\theta + 2\abs{\cos\theta \cos(2\delta)}}{4} \nonumber \\
        & \ge 0\,,
\end{align}
with equality if and only if $\cos\theta=\cos(Jt/2)=0$.
Consequently, for an infinite chain the rate function of the Loschmidt echo takes the form:
\be
    r^\infty (t) = \ln 2 -\frac{1}{2} \ln \left [ 1 + \cos^2 \frac{Jt}{2} + 2 \abs{\cos \frac{Jt}{2} \cos \frac{J\Delta t}{2}} \right]\,. 
    \label{eq:r_infty}
\ee
%It is worth noting that, by comparison with the derivation in the Bell-state representation in Eq.~\eqref{eq:G_dimer}, 
%the role of the time-dependent eigenvalues of the transfer matrix in determining the Loschmidt amplitude becomes more transparent.

The nonanalyticities of the rate function originate from the switching of the dominant transfer-matrix
eigenvalue between $\lambda_+$ and $\lambda_-$~\cite{Heyl115,Sirker}. The critical times of the DQPTs
are therefore determined by the condition: $\abs{\lambda_+} = \abs{\lambda_-}$, which is equivalent to
\be
       \cos\left( \frac{Jt}{2} \right) \cos \left(\frac{J\Delta t}{2}\right) = 0\,.
       \label{eq:dqpt_condition} 
\ee
This condition gives two types of critical times:
\begin{itemize}
\item[(i)] $\cos(Jt/2)=0$, leading to $\Delta$-independent critical times 
    \be 
       t^*=(2m+1)\pi/J\qquad m=0,1,2,\cdots\,; 
       \label{eq:t}
     \ee
\item[(ii)] $\cos(J\Delta t/2)=0$, which produces $\Delta$-dependent critical times  
     \be 
       t_{\Delta}^*=(2m+1)\pi/\Delta J\qquad m=0,1,2,\cdots\,.
       \label{eq:t_delta}
     \ee
\end{itemize}
We note that at a $\Delta$-independent critical time, all eigenvalues have the same modulus: $\abs{\lambda_+}=\abs{\lambda_-}=\abs{\lambda_0}=1/2$
and the rate function reaches its maximum $r^\infty (t^*)=\ln 2$. 

%%%%%%%%%%%% FIG7 %%%%%%%%%%%%%%
\begin{figure}[t]
\includegraphics[width=8.3cm]{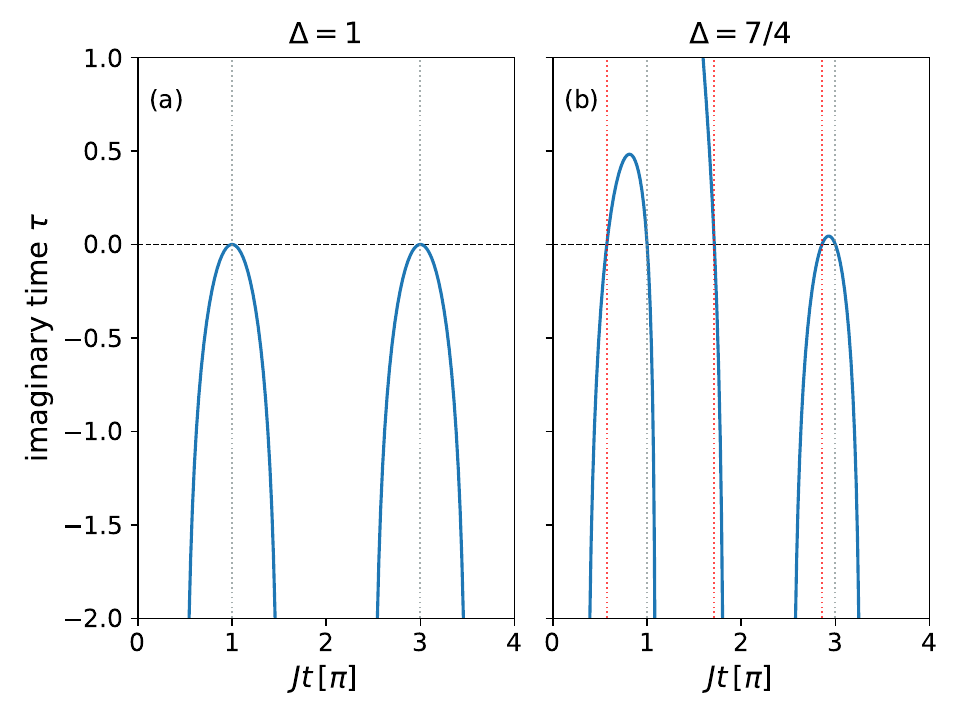}  % fisher_zeros.pdf
\caption{\label{fig:fisher}
Fisher zeros obtained by the condition $\abs{\lambda_+} = \abs{\lambda_-}$ in the complex time plane. (a) Isotropic case with $\Delta=1$: the curves only touches the real
time axis (horizontal dashed line) tangentially; (b) $\Delta=7/4$: a DQPT occurs when the Fisher-zero line crosses the real axis. Gray/red vertical dotted lines
mark the $\Delta$-independent/dependent critical times.}
\end{figure}

%%%%%%%%%%% FIG8 %%%%%%%%%%%

\begin{figure}[t]
\includegraphics[width=8.3cm]{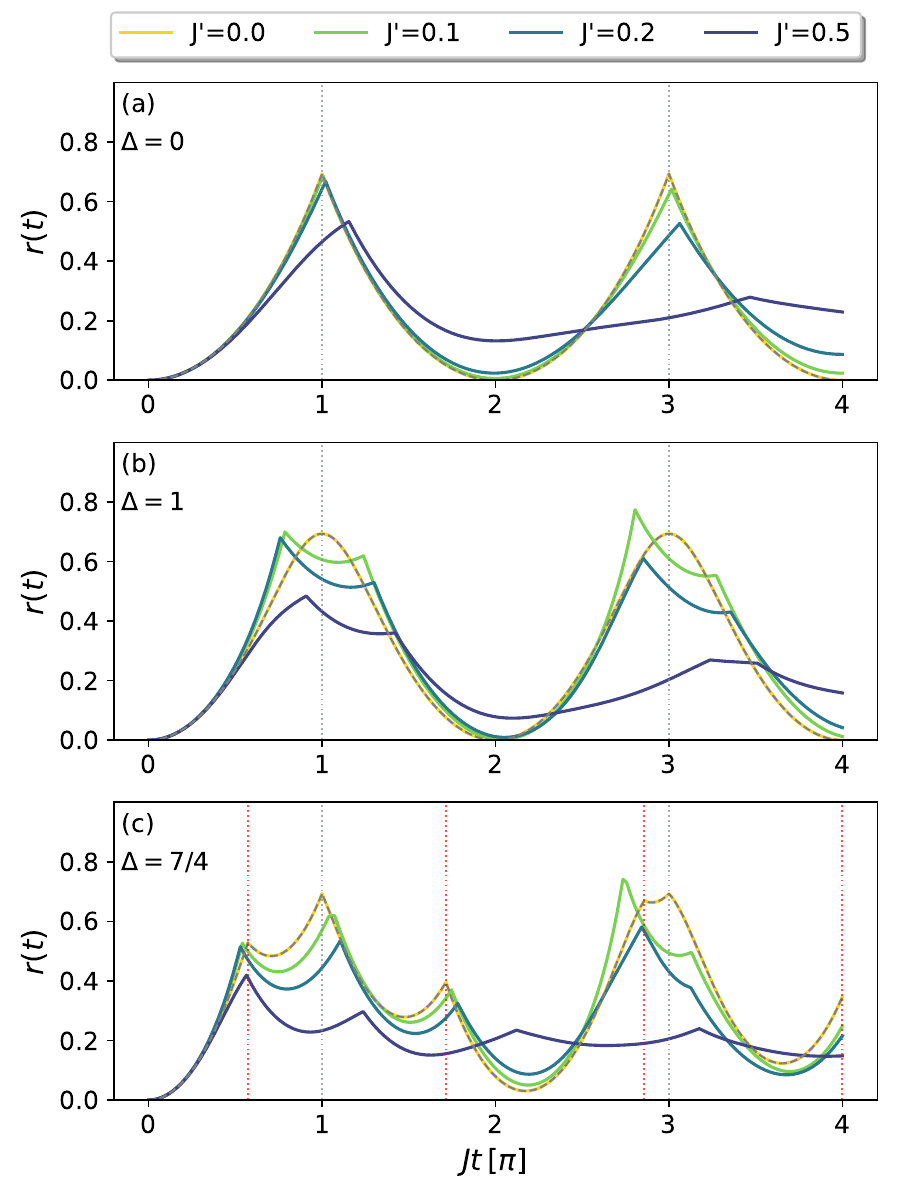}  % Je_return_h.pdf
\caption{\label{fig:r_finiteJ} ITEBD results for return rate functions with zero and finite odd couplings $J'$ for three values of
$\Delta$ in the thermodynamic limit. Gary dashed curves are plots of the analytical result in Eq.~\eqref{eq:r_infty}.
Gray/red vertical dotted lines mark the $\Delta$-independent/dependent critical times for $J'=0$.
The iTEBD simulations were performed with a maximum bond dimension $\chi_\text{max}=256$ or a singular value threshold $10^{-12}$, 
using a 2nd-order Trotter decomposition with a time step $dt=0.001\pi$.}
\end{figure}

Figure~\ref{fig:r_infty} presents the exact rate function $r^\infty(t)$ for various values of $\Delta$ in the thermodynamic limit,
overlaid with exact solutions for several finite systems. 
In general, the $\Delta$-dependent contributions give rise to a rich structure in the rate function.
%There are kink-like $r(t)$ structures observed in some finite size cases that disappear in the $n\to \infty$ limit
%and Loschmidt zeros observed in particular system sizes revival in the  infinite chain. 
There are two cases without the $\Delta$-dependent critical times: $\Delta=0$ and $\Delta=1$.
For $\Delta=0$, the rate function simplifies to $r_{\Delta=0}(t)=\ln 2- \ln[1+\abs{\cos(Jt/2)}]$ and develops kinks at $r^*$. 
For the isotropic case with $\Delta=1$, the rate function becomes a smooth function $r_{\Delta=1}(t)=\ln 2 - (1/2)\ln [1 +3\cos^2(Jt/2) ]$,
exhibiting maxima at $t_\Delta^*=(2m+1)\pi/J$ with 
$r_{\Delta=1}(t_\Delta^*)=\ln(2)$. The absence of nonanalyticities in the return rate function in the thermodynamic limit
indicates that, within the quench protocol considered here, no DQPTs occur in the isotropic XXX chain.
%, although there are kink-like structures observed in finite-size rate functions [see Figs.~\ref{fig:return_even}(c) and \ref{fig:return_odd}(c)]. 

Further insight into the absence of DQPTs at $\Delta=1$ can be obtained by analyzing the Fisher zeros through analytic continuation of the Loschmidt
amplitude to complex time $z = t + i\tau$~\cite{Heyl,Heyl_rev,Sirker}. Fisher zeros in the complex plane are determined by the condition $\abs{\lambda_+} = \abs{\lambda_-}$ 
as a function of $z$. 
For $\Delta=1$, we obtain a closed form for the corresponding curves of Fisher zeros::
\be
     \tau(t) = \ln[ -\cos(Jt) ],\qquad \cos(Jt) < 0\,,
     \label{eq:fisher_XXX} 
\ee
which restricts $Jt$ to intervals such as $(\pi/2, 3\pi/2) + 2\pi k$ for $k=0, 1, 2, \cdots$.
As shown in Fig.~\ref{fig:fisher}~(a),
The lines of Fisher zeros described by Eq.~\eqref{eq:fisher_XXX} for $\Delta=1$ 
lie entirely in the lower half-plane and only touches the real-time axis tangentially at the would-be critical times $t^*=(2m+1)\pi/J$.
%as shown in Fig.~\ref{fig:fisher}(a).
This behavior contrasts with the generic case $\Delta \neq 1$, 
where the Fisher-zero lines intersect the real-time axis at $t^*$ [see  Fig.~\ref{fig:fisher}~(b)],
leading to nonanalyticities in the rate function.
The absence of a DQPT at the XXX point for this particular quench is accidental: the two independent frequencies in Eq.~\eqref{eq:dqpt_condition}, 
$Jt/2$ and $J\Delta t/2$, become identical at $\Delta=1$. As a result, the two families of DQPTs coincide and effectively cancel each other.

The DQPTs in the fully dimerized XXZ chain considered here are symmetry-protected by 
the $\mathbb{Z}_2 \times \mathbb{Z}_2$ symmetry that distinguishes the two dimerized phases~\cite{Tasaki}.
These transitions are robust against adding finite couplings $J'$ on the odd links, provided $J'< J$~\cite{Heyl115,Ueda,Heyl_rev}.
As shown in previous studies for $\Delta=0$~\cite{Heyl115}, such perturbations merely shift the critical times in the short- and intermediate-time regimes. 

In Fig.~\ref{fig:r_finiteJ}, we present numerical results for finite $J'$ using the TeNPy library~\cite{Tenpy} in the framework of
the infinite-time evolving block decimation (iTEBD) algorithm~\cite{Vidal},
which further demonstrate the robustness of the DQPTs discussed in this work.
We note that introducing $J'>0$ ($J'< J$) restores DQPTs at the $\Delta=1$ point, since the condition in Eq.~\eqref{eq:dqpt_condition} no longer applies
and DQPTs reemerge as a generic consequence for a quench between two topologically distinct phases. 
At this special case, the restored DQPTs are characterized by double-kink structures (rather than one kink) in the rate function around $t^*$.
This feature may be attributed to the lifting of degeneracy in the transfer-matrix spectrum through the finite coupling $J'$.

\subsection{Finite-size scaling}
\label{sec:fss}
The transfer matrix method yields the exact Loschmidt amplitude in Eq.~\eqref{eq:G_finite} for arbitrary finite $n$,
in full agreement with the results obtained in the Bell-state representation.

Based on the exact results obtained here, we identify a universal $J\Delta$-independent finite-size scaling rule for 
the Loschmidt echo $\mathcal{L}^*$ and the return rate function $r^*$
at the $\Delta$-independent critical times $t^*=(2m+1)\pi/J$ for odd $n$ and all values of $J\Delta$:
\be
     {\cal L}^* = (1/2)^{2(n-1)},\; r^*=(1-1/n)\ln 2\,.
\ee 
This arises from the cancellation of $\lambda^n_+$ and $\lambda^n_-$ for odd $n$ at $t^*$, 
leaving only ${\cal L}^*=\abs{2\lambda_0}^2 =\abs{i^n(-1)^{mn}/2^{n-1}}^2$.

For even $n$, exact zeros of the Loschmidt echo can occur at finite system sizes, as noted in Sec.~\ref{sec:echo}.
However, no universal pattern governing these zeros has been identified, reflecting the nonmonotonic behavior of $\mathcal{L}$ in finite systems.
Below, we list several representative cases, assuming $J=1$:
\begin{itemize}
\item For $\Delta=0$, we identify $\mathcal{L}^*=0$ at $t^*$ for $n=2 \pmod 4$; %$n=4\nu+2, \nu=0,1,2,\cdots$;
\item For $\Delta=1/2$, $\mathcal{L}^*=0$ occurs at $t^*$ for $n=4 \pmod 8$ at $t_\Delta^*$ for $n=2 \pmod 4$; 
\item For $\Delta=7/4$, $\mathcal{L}^*=0$ occurs at $t^*$ for $n=8(2\nu+1)$ with $\nu=0,1,2,\cdots$;
\item For $\Delta=1$, $\mathcal{L}>0$ for all real $t$, i.e. no Loschmidt zeros are found.
\end{itemize}

\section{Digital quantum simulation}
\label{sec:simulation}

In this section we present the relevant setups for our simulations using the IBM Quantum Platform, %computer {\it ibm\_fez}, 
and show simulation results for time-dependent entanglement entropies and the Loschmidt echo.

The simulation data presented here were obtained from the quantum device {\it ibm\_fez}, a 156 qubit-system of type Heron, 
and a few from a noisy simulator based on the real device.

%\subsection{Amplitude estimation}
\subsection{Hadamard test for $a_k$ estimation}
\label{sec:hadamard}

%%%%%%%%%%% FIG9 %%%%%%%%%%%

\begin{figure}[t]
\includegraphics[width=8.3cm]{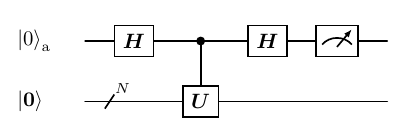}  % HT_circuit.pdf
\caption{\label{fig:HT} Hadamard test circuit for computing  the real part of the expectation value $\text{Re}(\trip{\bm{0}}{\bm{U}}{\bm{0}})$, 
where $\ket{\bm{0}}=\ket{0}^{\otimes N}$ is an $N$-qubit state 
and $\ket{0}_a$ is the initial state of an ancilla qubit.} 
%In our simulation, we set $\bm{U}=\bm{U}^\dag_{\Phi_k} \bm{U}_{\Psi_0}$, $k\in k^*$ to obtain 
%$a_k=\skp{\Phi_k}{\Psi_0}$, which are all real.}
\end{figure}

%%%%%%%%%%%%%%%%%%%%%%%%%%

%%%%%%%%%%% FIG10 %%%%%%%%%%%

\begin{figure}[t]
\includegraphics[width=8.6cm]{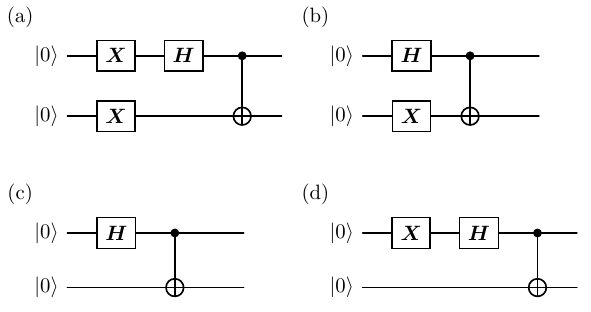} % bell.pdf
\caption{\label{fig:bell} Quantum circuits to generate four Bell states: (a) $\ket{\psi_0}=(\ket{01}-\ket{10})/\sqrt{2}$; 
(b) $\ket{\psi_1}=(\ket{01}+\ket{10})/\sqrt{2}$; (c)  $\ket{\psi_2}=(\ket{00}+\ket{11})/\sqrt{2}$; (d)  $\ket{\psi_3}=(\ket{00}-\ket{11})/\sqrt{2}$.}
\end{figure}

%%%%%%%%%%%%%%%%%%%%%%%%%%

%%%%%%%%%%% FIG11 %%%%%%%%%%%

\begin{figure}[t]
\includegraphics[width=8.6cm]{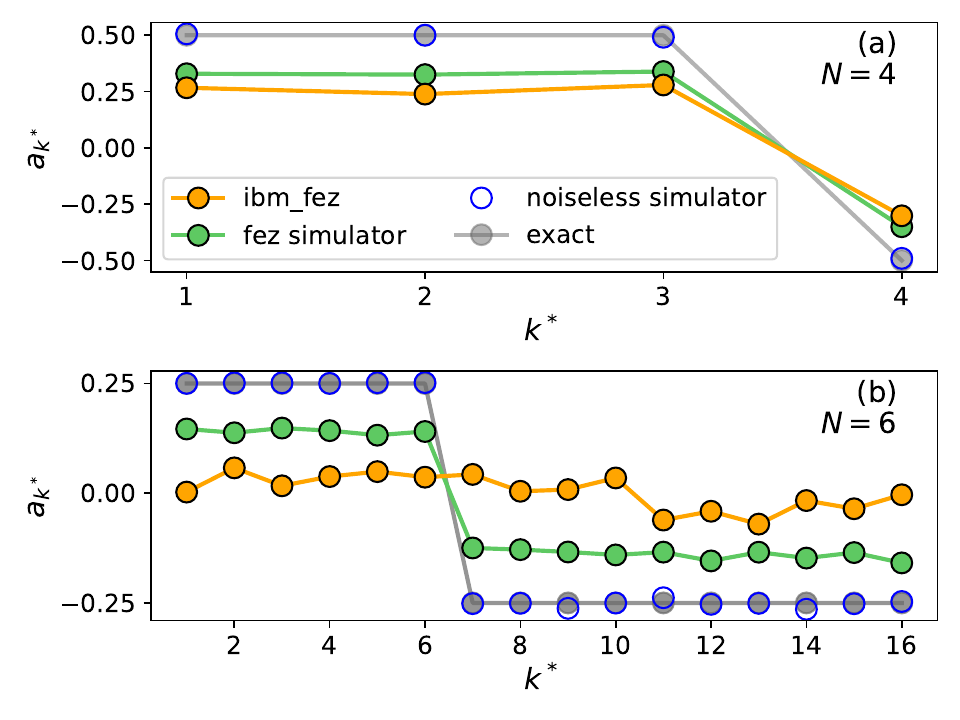} % coefficients_a.pdf
\caption{\label{fig:a_coef} Nonzero amplitudes $a_{k}=\skp{\Phi_k}{\Psi_0}$, $k\in k^*$ for chain lengths $N=4$ (a) and $N=6$ (b), estimated using
the Hadamard test on the IBM quantum device {\it ibm\_fez}
and its noisy simulator, compared with results from noiseless simulations and exact solutions.}
\end{figure}

%%%%%%%%%%%%%%%%%%%%%%%%%%

Our first numerical experiment estimates the coefficients (amplitudes) $a_k=\skp{\Phi_k}{\Psi_0}$ using the Hadamard test~\cite{Hadamard}.
The Hadamard test circuit, shown in Fig.~\ref{fig:HT}, consists
of two Hadamard gates $\bm{H}$ acting on a single ancilla qubit,
which serves as the control qubit for a controlled-$\bm{U}$ gate.
The unitary is defined as $\bm{U}=\bm{U}_{\Phi_k}^\dag \bm{U}_{\Psi_0}$,
where $\bm{U}_{\Psi_0}$ and $\bm{U}_{\Phi_k}$ generate the states $\ket{\Psi_0}$ and $\ket{\Phi_k}$, respectively, 
from a register of $N$ qubits initialized to $\ket{\bm{0}}=\ket{0}^{\otimes N}$:
\be
    \ket{\Psi_0} = \bm{U}_{\Psi_0} \ket{\bm{0}},\qquad \ket{\Phi_k} = \bm{U}_{\Phi_k} \ket{\bm{0}}\,. 
\ee

%%%%%%%%%%% FIG12 %%%%%%%%%%%
%\onecolumngrid

\begin{figure*}[t!]
\includegraphics[width=16cm, clip]{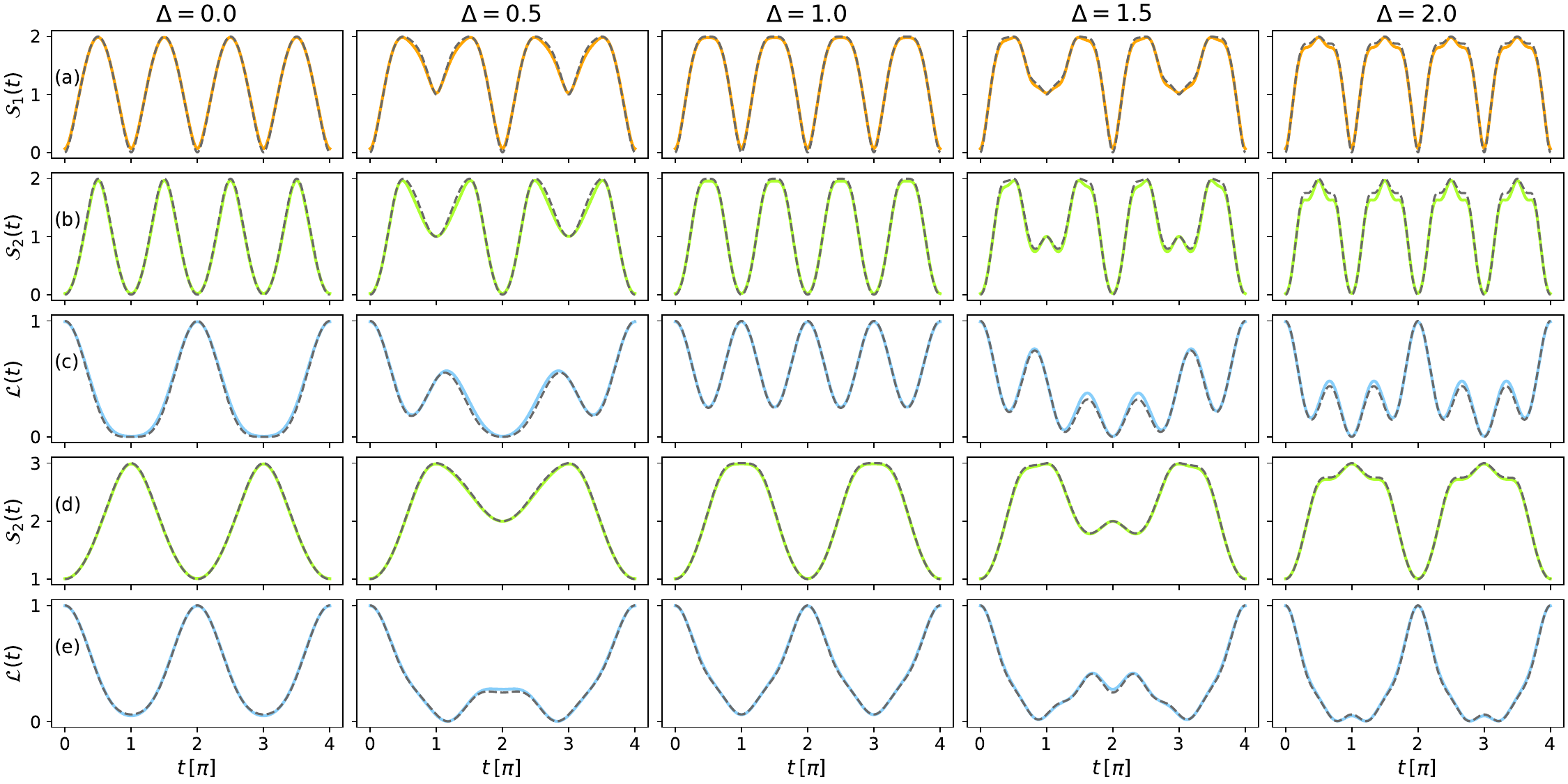} % HT_results_N4_N6.pdf
\caption{\label{fig:HT_N4N6}
Simulation results (solid lines) for the half-chain entanglement entropies and the Loschmidt echo
for a chain of length $N=4$ [Panel (a) -(c) ] and $N=6$ [Panel (d) and (e)] at various parameter $\Delta$, compared with
exact solutions (dashed lines). The simulation data are based on the normalized coefficients $a_k$, estimated using the Hadamard test on
the quantum device {\it ibm\_fez} for $N=4$ and on the noisy ibm simulator for $N=6$.}
\end{figure*}

%\twocolumngrid
%%%%%%%%%%%%%%%%%%%%%%%%%%

Since $\ket{\Psi_0}$ and $\ket{\Phi_k}$, are both product states of Bell
states, they can be constructed via tensor products of following unitaries
[Fig.~\ref{fig:bell}]:
\begin{align}
            \bm{U}_{\psi_0} &= \bm{C_X} (\bm{H}\otimes \bm{I}) (\bm{X}\otimes\bm{X})\\ 
            \bm{U}_{\psi_1} &= \bm{C_X} (\bm{H}\otimes \bm{X})\\
	    \bm{U}_{\psi_2} &= \bm{C_X} (\bm{H}\otimes \bm{I}) \\
	    \bm{U}_{\psi_3} &= \bm{C_X} (\bm{H}\otimes \bm{I}) (\bm{X}\otimes\bm{I})\,,	
\end{align}
where $\bm{C_X}$ denotes the CNOT gate, $\bm{X}$ the Pauli-$X$ gate, and $\bm{I}$ represents the identity operator.

To implement the controlled-$\bm{U}$ gate, we first construct the circuits preparing  $\ket{\Psi_0}$ and $\ket{\Phi_k}$
and convert them into unitary operators using the {\tt to\_gate()} function in Qiskit~\cite{Qiskit}.
We then form $\bm{U}=\bm{U}_{\Phi_k}^\dag \bm{U}_{\Psi_0}$ and convert it to a controlled operation,
\be
\bm{C}_{\bm{U}} = \begin{pmatrix}
\bm{I} & 0 \\
0 & \bm{U}
\end{pmatrix}\,,
\ee
using {\tt ControlledGate}, with an ancilla qubit as the control qubit initialized by a Hadamard gate.

The ancilla qubit is measured in the $Z$ basis after a second Hadamard gate.
The probabilities of measuring $0$ and $1$ in the ancilla qubit are
%\be
%  \begin{split}
\begin{align}
     %P_0 &= \frac{1}{4} \trip{\bm{0}}{(\bm{I}+\bm{U}^\dag)(\bm{I}+\bm{U} ) }{\bm{0}}\,, \\
     %P_1 & = \frac{1}{4} \trip{\bm{0}}{(\bm{I}-\bm{U}^\dag)(\bm{I}-\bm{U} ) }{\bm{0}}\,. 
     P_0 &= \frac{1}{4} \left( 2 + \trip{\bm{0}}{\bm{U} + \bm{U}^\dag}{\bm{0}} \right)\,, \nonumber \\
     P_1 & = \frac{1}{4} \left( 2 - \trip{\bm{0}}{\bm{U} + \bm{U}^\dag}{\bm{0}} \right)\,.  
\end{align}
%  \end{split}
%\ee
Thus, we obtain the coefficient $a_k\in\mathbb{R}$, the real part of the amplitude, by
\be
     a_k = \text{Re}\left[ \trip{\bm{0}}{\bm{U}}{\bm{0}}  \right] = P_0-P_1\,.
\ee

Based on the selection rules described in Sec.~\ref{sec:a_k}, we apply the
Hadamard test exclusively to those active states $\ket{\Phi_{k^*}}$ that yield
nonzero $a_{k^*}$.  Results for $N=4$ and $N=6$, simulated on the IBM device
{\sl ibm\_fez} and its noisy simulator, are shown in Fig.~\ref{fig:a_coef}.
For each data point, $N_M=2^{13}$ measurements for $N=4$ and
$N_M=2^{14}$ for $N=6$ were performed.   
For $N=4$, only four active states contribute to the superposition: three with positive
coefficients and one with a negative coefficient.  The Hadamard test correctly
reproduces the signs of these coefficients, though the measured magnitudes are
systematically smaller than the exact values.  For $N=6$, the number of active
states increases to 16, and the accuracy of results from the real device
declines significantly.  This accuracy decline stems from two factors: the
rapidly increasing circuit depth (exceeding $300$ for $N=6$) and the decreasing
amplitude magnitudes ($\abs{a_{k^*}}=0.25$ for $N=6$) for longer chains.

%%%%%%%%%%% FIG13 %%%%%%%%%%%

\begin{figure}[t]
\includegraphics[width=8.3cm, clip]{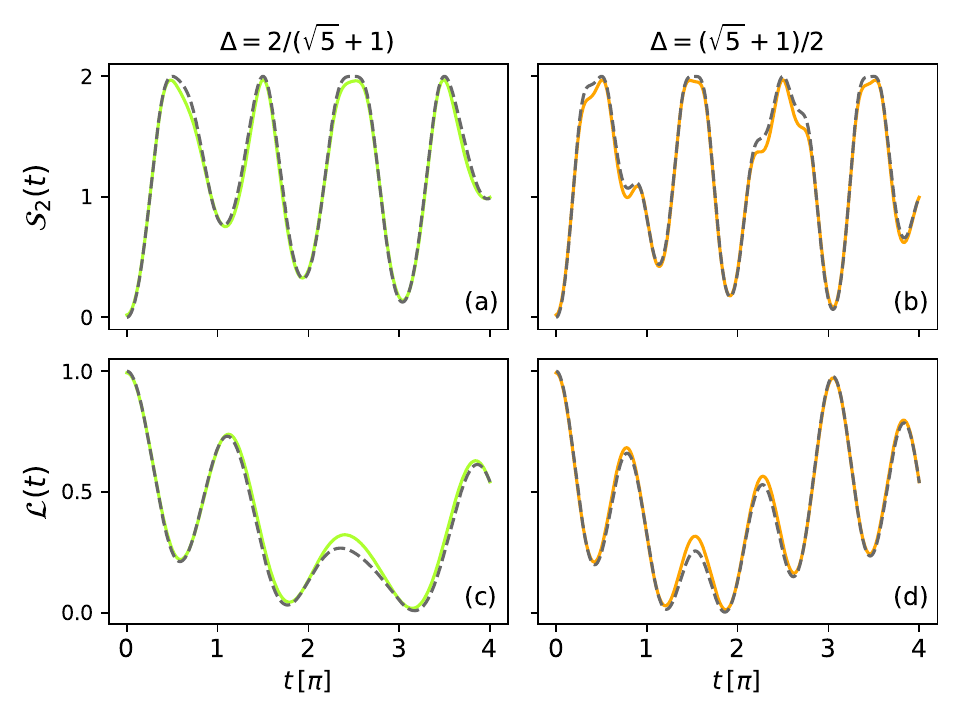} % HT_data_D0618.pdf
\vskip-2mm
\caption{\label{fig:HT_N4_D0618} Simulation results (solid lines) for the half-chain R\'enyi entanglement entropy and the Loschmidt echo 
for a chain of $N=4$ with $\Delta=2/(\sqrt{5}+1)$ and $\Delta=(\sqrt{5}+1)/2$. The simulation data are based
on the normalized coefficients $a_k$, estimated using the Hadamard test on {\it ibm\_fez}.}
\end{figure}

%%%%%%%%%%%%%%%%%%%%%%%%%%

We constructed the time-evolved state $\ket{\Psi_0(t)}$ using these estimated
amplitudes and renormalized them according to the condition:
\be
     a_{k^*}\; \leftarrow\; \frac{a_{k^*}}{\sqrt{\sum_{k^*} a_{k^*}^2}} 
\ee
This normalization procedure serves as an error mitigation strategy since the
exact coefficients $a_{k^*}$ inherently satisfy this normalization condition.
With the time-evolved state constructed, we then calculated the observables of
interest, including the von Neumann entanglement entropy of a half chain, 
the R\'enyi entanglement entropy, and the Loschmidt echo. 

Figure~\ref{fig:HT_N4N6} [panel (a)-(c)] presents the observable measurements for $N=4$ obtained using
the normalized amplitudes $a_k$ from the Hadamard test on the IBM device {\sl ibm\_fez}.
For $N=6$, however, the inaccurate coefficients estimated from the real device produce physically
meaningless results for the observables. Consequently, we employed
coefficients obtained from the noisy simulator to calculate the observables
shown in Fig.~\ref{fig:HT_N4N6} (d) and (e).

The entanglement entropy and the Loschmidt echo produced by the estimated amplitudes from
{\sl ibm\_fez} for the $N=4$ chain with an irrational $\Delta=2/(\sqrt{5}+1)$ are presented in 
Fig.~\ref{fig:HT_N4_D0618}. The data are also in good agreement with the exact solutions.

\subsection{Time evolution circuit}
\label{sec:t-circuit}

%%%%%%%%%%% FIG14 %%%%%%%%%%%

\begin{figure}[t]
\includegraphics[width=8cm, clip]{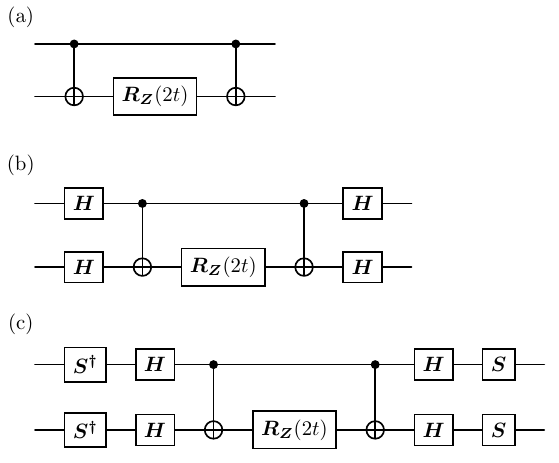} % expU.pdf 
\vskip-2mm
\caption{\label{fig:expU} Quantum circuits to implement the time-evolution operators: 
(a) $\exp(-it \bm{Z} \bm{Z})$; (b) $\exp(-it \bm{X} \bm{X})$; (c) $\exp(-it \bm{Y} \bm{Y})$.
}
\end{figure}

%%%%%%%%%%%%%%%%%%%%%%%%%%

While the Hadamard test enables direct amplitude estimation of time-evolved
states in the Bell basis, it yields accurate results 
only for very small systems on real quantum devices. As system size grows, 
circuit depth increases substantially, degrading accuracy on noisy hardware.
Error mitigation techniques, such as readout error mitigation, zero-noise extrapolation, and Pauli twirling,
can partially compensate for hardware noise~\cite{Mitig_rev}; however, the exponential decay of amplitude magnitudes with system size
ultimately limits the practical scalability of this approach in our simulations.

A more scalable approach is to construct time-evolved states by applying
time-evolution operators to the initial state. The key challenge in
simulating general quantum many-body dynamics on digital quantum devices is
the design of shallow circuits for time evolution that avoid substantial
Trotter errors~\cite{Trotter}. For the fully dimerized model considered here, this challenge
is overcome: the time-evolution circuit naturally decomposes into a sequence
of decoupled two-qubit gates and requires no Trotter approximation, resulting
in a constant circuit depth independent of evolution time $t$.

The time evolution operator governed by $\mathcal{H}_2$ can be decomposed as
\be
      U(t,\Delta) = \bigotimes_j U_{2j}(t,\Delta)\,,
\ee
with
\begin{align}
  U_{2j}(t,\Delta) & =\exp\biggl[ -\frac{it}{4} (\bm{X}_{2j}\bm{X}_{2j+1} +\bm{Y}_{2j}\bm{Y}_{2j+1} +\Delta \bm{Z}_{2j}\bm{Z}_{2j+1})\biggr]
                    \nonumber \\
                   & =\exp(-\frac{it}{4} \bm{X}_{2j}\bm{X}_{2j+1})\exp(-\frac{it}{4} \bm{Y}_{2j}\bm{Y}_{2j+1}) \cdot \nonumber \\
                   &\qquad \exp(-\frac{it}{4}\Delta \bm{Z}_{2j}\bm{Z}_{2j+1})\,,
\end{align}
in terms of Pauli gates $\bm{X}$, $\bm{Y}$ and $\bm{Z}$.
The quantum circuits implementing the three exponential operators with $ZZ$, $XX$, and $YY$ couplings
are illustrated in Fig.~\ref{fig:expU}(a), \ref{fig:expU}(b), and \ref{fig:expU}(c), respectively.
In this setup, the exponential operators are related to each other through the basis transformation:
$({\bm{u}^X})^\dag\bm{Z}\bm{u}^X=\bm{X}$ and $({\bm{u}^Y})^\dag\bm{Z}\bm{u}^Y=\bm{Y}$ with $\bm{u}^X=\bm{H}$
and $\bm{u}^Y=\bm{HS}^\dag$, %where the adjoint of the phase gate $\bm{S}$ is used. 
where $\bm{S}$ is a phase shift gate which rotates a state vector a $\pi/2$ radian around the $Z$-axis.

\subsection{Randomized measurements}

To probe the properties of the time-evolved quantum state prepared by the time-evolution circuit, we employ a random Pauli
measurement scheme, in which each qubit is measured in a randomly chosen $X$,
$Y$ or $Z$ basis~\cite{Shadow,RM_rev}.  The local random unitaries used to
rotate the measurement basis can be sampled from the single-qubit Clifford
ensemble.

%%%%%%%%%%% FIG15 %%%%%%%%%%%

\begin{figure}[t!]
\includegraphics[width=8.6cm, clip]{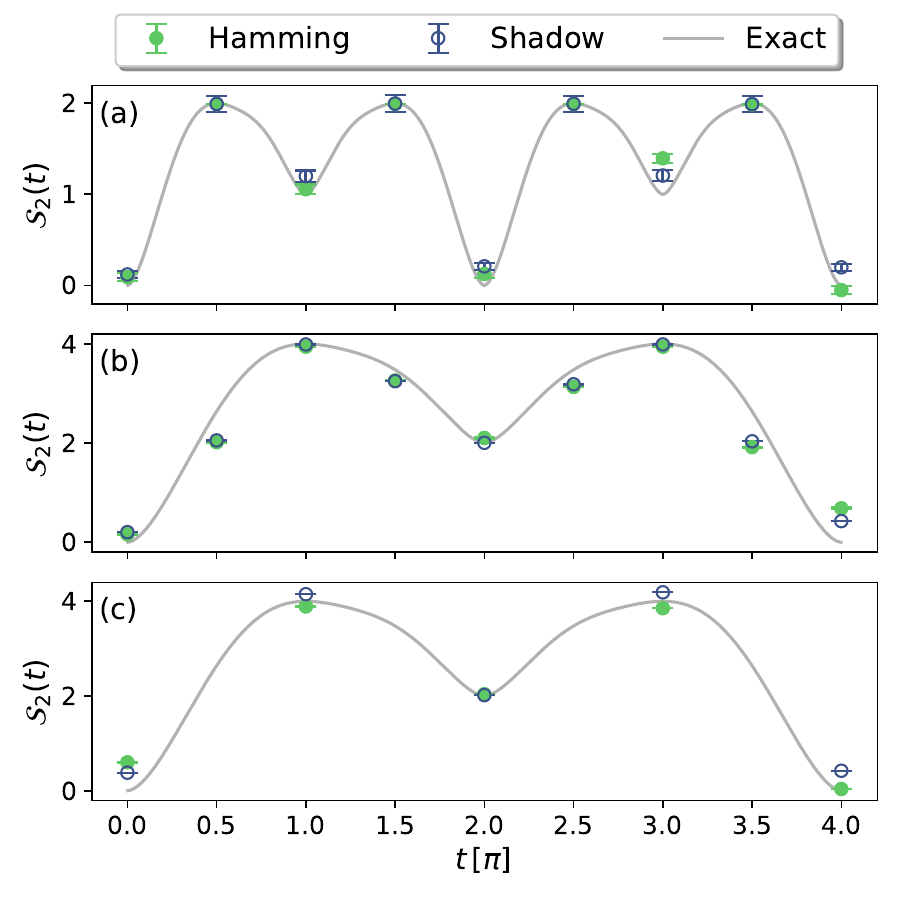} % RM_S_h.pdf
\caption{\label{fig:RM_S} Results of random Pauli measurements for dynamical half-chain R\'enyi entanglement entropy of the XXZ chain with $\Delta=0.5$ and 
with $N=4$ (a), $N=8$ (b), and $N=12$ (c) sites. The same measurement outcomes are analyzed using Hamming distance method [Eq.~\eqref{eq:rm_purity}] and
classical shadow technique [Eq.~\eqref{eq:shadow_purity}], and compared with the exact solutions (solid lines).
For $N=8$, we consider the subsystem consisting of the four bulk qubits in the open chain, while for 
$N=4$ and $N=12$ the subsystems consist of $N/2$ adjacent qubits in a chain with PBC.}
\end{figure}

%%%%%%%%%%%%%%%%%%%%%%%%%%

The measurement protocol proceeds as follows:
(i) Prepare the postquench state $\rho(t)$ of $N$ qubits at time $t$ using the time-evolution circuit.
(ii) Apply a random unitary transformation, $\rho \rightarrow \bm{U}\rho \bm{U}^\dag$, 
where $\bm{U}=\bigotimes_{j=1}^N \bm{u}_j$ and each $\bm{u}_j$ is randomly drawn from 
$\{ \bm{H}, \bm{H}\bm{S}^\dag,\bm{I} \}$, thereby rotating the local measurement basis to a randomly chosen Pauli basis
$\{X, Y, Z \}$.
(iii) Measure each qubit in the standard $Z$ basis, obtaining a bitstring $\mathbf{b}=z_1\cdots z_N$ with  $z_j=0,1$. 
For each fixed unitary $\bm{U}$, the measurement is repeated $N_M$ times (taking $N_M$ shots). The full procedure is then repeated for $N_U$
independently sampled unitaries. 
The resulting $N_U\times N_M$ bitstrings and corresponding sets of single-qubit unitaries are stored for classical post-processing.

We apply two post-processing protocols to estimate the second-order R\'enyi entanglement entropy (EE) and 
the Loschmidt echo. In the first protocol, the purity of a subsystem $A$, required for the R\'enyi EE, 
is estimated from statistical correlations of randomized measurements via~\cite{RM_PRA97,RM,RM_PRA}:
\be
%     \text{Tr}[\rho_A^2] = \frac{2^{N_A}}{N_U} \sum_{\bm{U}} \sum_{\mathbf{b}_A, \mathbf{b}'_A} (-2)^{-D[\mathbf{b}_A, \mathbf{b}'_A]} P_{\bm{U}} (\mathbf{b}_A) P_{\bm{U}} (\mathbf{b}'_A)\,,
\text{Tr}[\rho_A^2] = \frac{2^{N_A}}{N_U} \sum_{m} \sum_{\mathbf{b}_A, \mathbf{b}'_A} (-2)^{-D[\mathbf{b}_A, \mathbf{b}'_A]} P^{(m)}(\mathbf{b}_A) P^{(m)}(\mathbf{b}'_A)\,,
\label{eq:rm_purity}
\ee
where $D[\mathbf{b}_A, \mathbf{b}'_A] \equiv \abs{\{j\in A \mid z_j \neq z'_j \}}$ is the Hamming distance between two bitstrings $\mathbf{b}_A=z_1\cdots z_{N_A}$ and $\mathbf{b}'_A=z'_1\cdots z'_{N_A}$, and $P^{(m)}(\mathbf{b}_A)$ denotes the probability of measuring bitstring $\mathbf{b}_A$ under the $m$th Pauli measurement setting $\bm{U}^{(m)}$.  
Similarly, the overlap between two $N$-qubit states $\rho$ and $\rho'$ can be estimated using~\cite{RM_PRA}:
\be
     %\text{Tr}[\rho_1 \rho_2] = \frac{2^{N}}{N_U} \sum_{\bm{U}} \sum_{\mathbf{b}, \mathbf{b}'}  (-2)^{-D[\mathbf{b}, \mathbf{b}']} P^{(1)}_{\bm{U}} (\mathbf{b}) P^{(2)}_{\bm{U}} (\mathbf{b}')\,,
     \text{Tr}[\rho \rho'] = \frac{2^{N}}{N_U} \sum_{m} \sum_{\mathbf{b}, \mathbf{b}'}  (-2)^{-D[\mathbf{b}, \mathbf{b}']} P^{(m)} (\mathbf{b}) P'^{(m)} (\mathbf{b}')\,,
\label{eq:rm_overlap}
\ee
where $P^{(m)}$ and $P'^{(m)}$ are the probabilities 
based on $\rho$ and $\rho’$, respectively.
From the subsystem purity, we compute the R\'enyi EE of a half-chain, 
while identifying $\rho = \rho(0) =\ket{\Psi_0}\bra{\Psi_0}$ and $\rho' = \rho(t)$ yields the Loschmidt echo.

%%%%%%%%%%% FIG16 %%%%%%%%%%%

\begin{figure}[t!]
\includegraphics[width=8.6cm, clip]{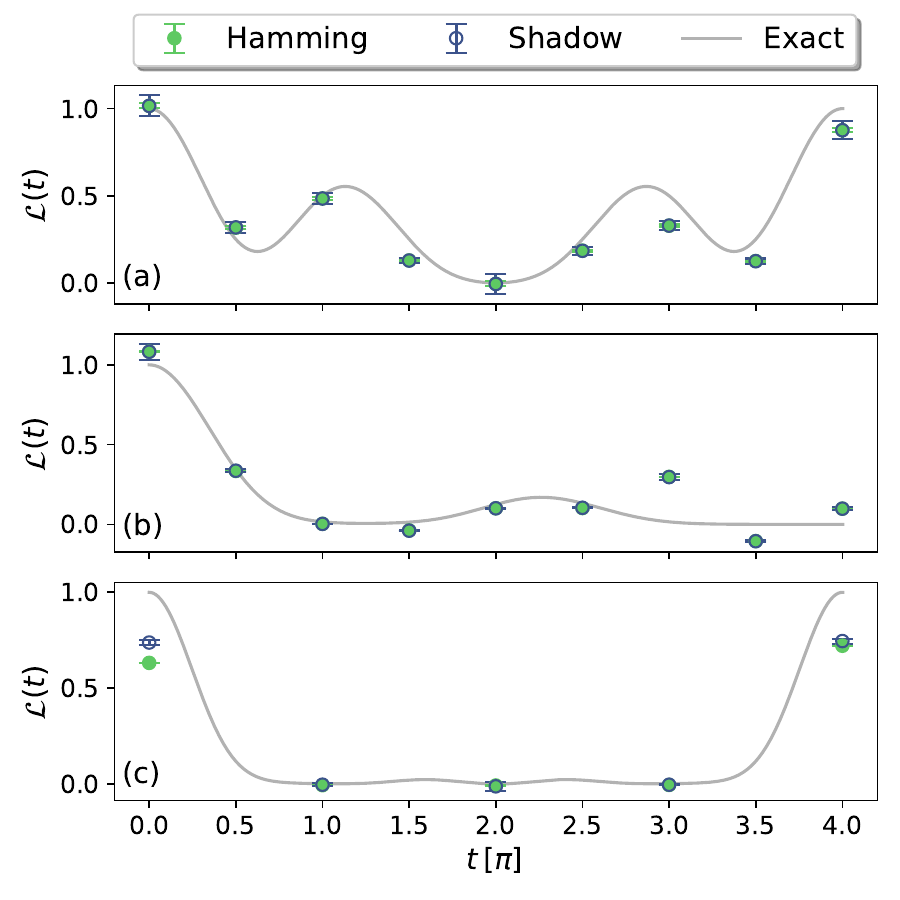} % RM_echo_h.pdf
\caption{\label{fig:RM_echo} Results of random Pauli measurements for the Loschmidt echo in the XXZ chain with $\Delta=0.5$ and
with $N=4$ (a), $N=8$ (b), and $N=12$ (c) sites. For $N=8$, the system has open boundary conditions, while for $N=4$ and $N=12$, periodic 
boundary conditions are considered. The results were evaluated using the same measurement data as in Fig.~\ref{fig:RM_S}.}
\end{figure}

%%%%%%%%%%%%%%%%%%%%%%%%%%

A second post-processing protocol employs the pairs of random unitaries and measurement outcomes, $\{(\bm{U}, \mathbf{b} ) \}$, 
to construct classical shadows of the quantum state ~\cite{Shadow,RM_rev}.
If $\ket{\mathbf{b}}=\ket{z_1\cdots z_N}$ is the
measurement outcome in the $Z$ basis after applying the unitary transformation
$\bm{U}=\bigotimes_{j=1}^N \bm{u}_j$ to an $N$-qubit state $\rho$, a classical snapshot of the state can be constructed as
$\bigotimes_{j=1}^N \bm{u}_j^\dag \ket{z_j}\bra{z_j} \bm{u}_j$.
The reduced state on a subsystem $A$ is then reconstructed via 
applying the single-qubit inverse of the measurement channel on each qubit in  $A$~\cite{Shadow, RM_rev}:
\be
      \hat{\rho}_A(\bm{U})
      =\bigotimes_{j \in A} \biggl[ 3 \bm{u}_j^\dag \ket{z_j}\bra{z_j} \bm{u}_j -\bm{I}  \biggr]\,.
\label{eq:shadow}
\ee
%Given $N_M$ measurement outcomes in setting $\bm{U}^{(m)}$, 
%
The collection of inverted snapshots, $\{\hat{\rho}_A^{(m)} \}_{m=1}^{N_U}$, %obtained over $N_U$ measurements, 
constitutes the classical shadow of the state, where $\hat{\rho}_A^{(m)}$ is the shot average
\be
     \hat{\rho}_A^{(m)} = \frac{1}{N_M} \sum_{k=1}^{N_M} \hat{\rho}_{A,k}(\bm{U}^{(m)})
     \label{eq:shot_av}
\ee
in setting $\bm{U}^{(m)}$. 
%In our implementation, the classical shadows are built using the same measurement data as
%in the first protocol based on weighted averages of Hamming distances [see Eq.~\eqref{eq:rm_purity}].
%That is, we have $M=N_U \times N_M$.

In this protocol, the subsystem purity is estimated as an average over all pairs of shadows~\cite{Shadow, RM_rev}:
\be
   \text{Tr}[\rho^2_A] = \frac{1}{N_U(N_U-1)} \sum_{m \neq m'} \text{Tr} [\hat{\rho}_A^{(m)} \hat{\rho}_A^{(m')}]\,,
   \label{eq:shadow_purity}
   %\quad m, m'=1,\cdots, M\,.
\ee
for $m,\,m'=1,\cdots, N_U$. To obtain the Loschmidt echo, we use the exact density matrix  $\rho(0)=\ket{\Psi_0}\bra{\Psi_0}$ of the initial state $\ket{\Psi_0}$ to evaluate
the overlap via the ensemble average:
\be
     \mathcal{L}(t) = \frac{1}{N_U} \sum_{m} \text{Tr} \left [\rho(0) \hat{\rho}^{(m)}(t) \right ]
\ee

The classical representation obtained from random Pauli measurements enables the estimation of a wide range of observables in the stage of post-processing.
In addition to the R\'enyi EE and the Loschmidt echo, here we also evaluate an order parameter defined as the expectation value of the twist operator~\cite{Lieb,Todo},
\be
              U_z = \exp\left(i \frac{\pi}{N} \sum_{j=1}^N j \bm{Z}_j \right)\,.
              \label{eq:Uz}
\ee
This twist order parameter distinguishes between two symmetry-protected topological phases, 
taking the value $\e{U_z}= [\cos(\pi/N) ]^{N/2}$ in the trivial (odd-dimer) phase and $\e{U_z}=-[\cos(\pi/N) ]^{N/2}$ in the topological  (even-dimer) phase~\cite{Todo}.
In Ref.~\cite{XX}, this quantity was estimated via direct measurements in the computational basis, where the twist operator is diagonal.
Here, instead, we infer it from random Pauli measurement outcomes using the same dataset as for the R\'enyi EE and the
Loschmidt echo, thus requiring no additional measurements on the quantum device.

Within the classical shadows framework, the twist order parameter $\e{U_z}=\text{Tr}[U_z \rho]$ is formally given by
\be
     \e{\hat{U}_z}= \sum_m \text{Tr} \left[ U_z \hat{\rho}^{(m)} \right]\,.
     \label{eq:eU}
\ee
To avoid expensive matrix operations, we instead rewrite Eq.~\eqref{eq:eU} by exploiting the algebra of Pauli operators.
%Exploiting the commutativity of Pauli operators on different sites, together with the generalized Euler formula 
In particular, using the generalized Euler formula
$\exp(i\Theta \bm{Z}) = \cos \Theta \bm{I} + i\sin \Theta \bm{Z}$,
we expand the exponential in Eq.~\eqref{eq:Uz} into a sum of identity and products of $\bm{Z}$ operators:
\be
           U_z = \sum_{\mathcal{P} \subseteq \{1, 2, \cdots, N \}} \alpha_\mathcal{P} \bm{O}_\mathcal{P}\,, 
          \label{eq:U_O}
\ee
where the operator string for subset $\mathcal{P}$ and its associated coefficient are given by
\be
       \bm{O}_\mathcal{P} = \bigotimes_{j=1}^N \bm{O}_j = 
       \bigotimes_{j=1}^N \begin{cases} 
       %\begin{cases}
              \bm{Z}_j & \text{if } j\in \mathcal{P} \\
              \bm{I}_j & \text{if } j\notin \mathcal{P}
        \end{cases}
\ee
\be
      \alpha_\mathcal{P} = \prod_j  \begin{cases}
       %\begin{cases}      
               i\sin \Theta_j  & \text{if } j\in \mathcal{P} \\  
               \cos \Theta_j                & \text{if } j\notin \mathcal{P}
      \end{cases}
\ee
with $\Theta_j = \pi j/N$. An unbiased estimator for the expectation value of the twist operator can then be constructed as
\be
   \e{\hat{U}_z} = \frac{1}{N_U} \sum_{m=1}^{N_U} \sum_{\mathcal{P} \subseteq \mathcal{Z}^{(m)}} \alpha_\mathcal{P}\cdot 3^{\abs{\mathcal{P}}} \eta^{(m)}_\mathcal{P}\,,
    \label{eq:twist_shadow}
\ee
where $\abs{\mathcal{P}}$ denotes the size of the subset $\mathcal{P}$, and  we define
\be
      \eta^{(m)}_\mathcal{P} = \frac{1}{N_M}\sum_{k=1}^{N_M} \prod_{j \in \mathcal{P}} (1-2z_j)\,,
      \label{eq:eta}
\ee
and
\be
      \mathcal{Z}^{(m)} = \{j \mid \bm{P}^{(m)}_j = \bm{Z}_j \}\,,
      \label{eq:Z_set}
\ee
where $\bm{P}^{(m)}_j \in \{\bm{X}, \bm{Y}, \bm{Z} \}$ is the local Pauli basis in setting $m$.
A derivation of this estimator is given in Appendix.~\ref{sec:Uz}.

%%%%%%%%%%% FIG17 %%%%%%%%%%%

\begin{figure}[t]
\includegraphics[width=8.6cm]{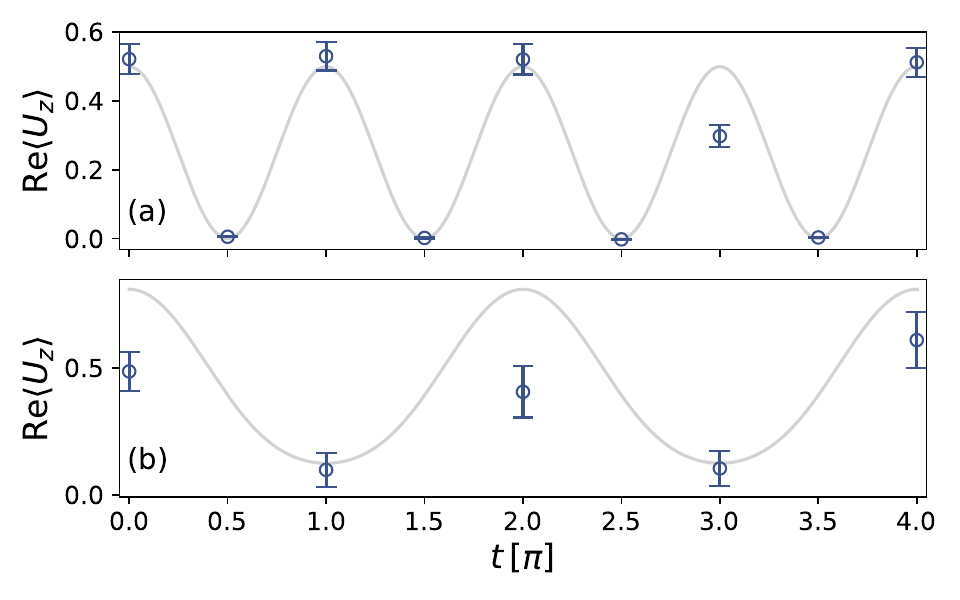} % Uz.pdf
\caption{\label{fig:twist} Estimating twist order parameter using classical shadows for XXZ chains with $\Delta=0.5$ and with $N=4$ (a) and $N=12$ (b) qubits.
The data collected from random Pauli measurements on the quantum device {\it ibm\_fez} are identical to those used in
Figs.~\ref{fig:RM_S} and \ref{fig:RM_echo} for both $N=4$ and $N=12$. The gray solid lines represent the corresponding exact results.} 
\end{figure}

%%%%%%%%%%%%%%%%%%%%%%%%%%

In Fig.~\ref{fig:RM_S} and Fig.~\ref{fig:RM_echo}, we present the results for the half-chain R\'enyi EE and the Loschmidt echo, respectively,
%comparing data obtained from the two post-processing protocols with the exact solutions.
comparing data obtained from random Pauli measurements with the exact solutions.
The same measurement data are post-processed using Hamming distance method and classical shadow technique.
According to previous studies~\cite{RM_PRA97,RM_PRA,Shadow,RM_rev}, for a fixed total number $M=N_M\times N_U$, the Hamming distance method yields more accurate estimates
when the number of measurement shots $N_M$ is large, whereas the classical shadow technique performs better when $N_U$ is large.
Without optimizing the ratio of $N_M$ and $N_U$ for our setting, we chose $N_U=2^6$ and $N_M=2^{13}$.

We consider systems with $\Delta=0.5$ and $N=4, 8$, and $12$ qubits. For $N=4$ and $N=12$, periodic boundary conditions (PBC) were implemented,
which require connectivity between the first and last qubits, typically achieved through ancilla qubits and additional gates.
For the $N=12$ system, we utilized the heavy-hexagon lattice structure of the IBM quantum processors to
select a ring of qubits along the hexagonal arrangement, allowing PBC to be realized
without extra gates or ancillas~\cite{XX}. For $N=8$, we simulated a chain with OBC to keep the circuit depth shallow for this system size.

Without any error mitigation, the simulated half-chain EE shows overall good agreement with the exact results.
For the Loschmidt echo, which involves the full system, noticeable deviations from the exact values appear at certain times.
Nonetheless, the two post-processing methods yield consistent results across all cases.
Moreover, compared to previous raw EE results for the XX case~\cite{XX}, 
the present results achieve significantly higher accuracy, which can be attributed to two main factors:
(i) the use of random Pauli measurements, which are more hardware-efficent than the Haar ranom unitaries employed in the earlier study, 
and (ii) substantial improvements in the IBM Quantum hardware.

Like the Loschmidt echo, the twist order parameter also involves contributions from the full system.
In Fig.~\ref{fig:twist}, we present results for this topological order parameter in
chains with PBC, obtained using the shadow estimator in
Eq.~\eqref{eq:twist_shadow}, alongside the corresponding exact results. For the
quench protocol considered here, the twist order parameter is purely real,
reflecting the inversion symmetry of the dynamical state at all times
$t$~\cite{XX}.  Furthermore, the twist order parameter remains positive
throughout the evolution, indicating that a system initialized in the odd-dimer
phase does not transition to the even-dimer phase after the quench. The
empirical results show satisfactory agreement with the exact values for the
short chain with $N=4$.  For the larger system with $N=12$, however, the
deviations from the exact results become pronounced, although the qualitative
behavior is still well captured.

\section{Summary and discussion}
\label{sec:summary}
We have analytically and numerically investigated quantum quenches in the XXZ chain with alternating interactions in the flat-band limit, 
with particular emphasis on the role of the anisotropy $\Delta$ in the $z$-component of the exchange coupling.  
The absence of dispersion in this limit enables an exact characterization of the non-equilibrium dynamics across the topological phase
boundary, providing valuable insight into the emergence of DQPTs.

By formulating the problem in the Bell basis, we established selection rules identifying all Bell states with nonzero expansion coefficients 
in the time-dependent state for arbitrary even system sizes. For a chain of length $N$, all nonzero coefficients have equal magnitude $\abs{a_k} = 2^{-(N/2-1)}$. This constrained structure greatly simplifies both analytical calculations and numerical verification.
We derived exact solutions for half-chain entanglement entropies and Loschmidt echoes in finite and infinite chains.
Both the entanglement entropies and the Loschmidt echoes
exhibit persistent oscillations, reflecting the absence of relaxation in this flat-band system. 
For $N>4$, these observables are periodic with period $2q\pi$ when $J\Delta=p/q$ is an irreducible fraction.

We identified two distinct classes of critical times associated with DQPTs.  The first class consists of $\Delta$-independent critical times
at $t^* = (2m+1)\pi/J$ with $m=0,1,2,\cdots$, while the second class arises from the anisotropy and leads to $\Delta$-dependent critical times
at $t_\Delta^* = (2m+1)\pi/J\Delta$. The coexistence of these two classes of critical times gives rise to a rich dynamical structure 
in the Loschmidt rate function. We demonstrated that DQPTs in the fully dimerized limit are robust against perturbations that include finite, 
weak couplings on the initially decoupled bonds. This robustness was verified numerically using iTEBD, which shows that the nonanalytic features 
of the rate function persist, albeit with shifted critical times, indicating that the observed DQPTs are not artifacts of the flat-band limit.

On the digital quantum simulation side, we performed two complementary experiments on IBM Quantum devices. 
First, we used the Hadamard test to directly evaluate scalar products between 
the initial state  $\ket{\Psi_0}$ and Bell basis vectors $\ket{\Phi_k}$, including their phase (sign) information.
Combined with the known energies of the Bell states, 
this procedure enables a full reconstruction of the time-evolved state in the Bell basis, from which dynamical observables can
be computed. By incorporating the normalization condition as a simple error mitigation strategy, 
this approach yields accurate results for small system sizes ($N=4$) on current hardware.
However, the exponential decay of amplitude magnitudes with increasing system size, together with the rapidly growing
circuit depth, makes this Hadamard-test approach impractical for $N>4$ in our simulations.

To overcome these limitations, we implemented time-evolution circuits that directly simulate the quench dynamics
and applied randomized measurement techniques. Due to the decoupled structure of the Hamiltonian in
the flat-band limit, the time-evolution gates can be realized exactly without Trotter decomposition, 
leading to intrinsically shallow circuits. We performed random Pauli measurements by sampling each
qubit in the $X$-, $Y$-, and $Z$-bases~\cite{Shadow}, and extracted dynamical observables
through classical post-processing using two equivalent frameworks --- the
Hamming-distance method~\cite{RM,RM_PRA97} and the classical shadow technique~\cite{Shadow,RM_rev}.
All observables, including the R\'enyi entanglement entropy, the Loschmidt echo,
and the twist order parameter, were obtained from the same measurement data,
highlighting the efficiency and versatility of the post-processing protocol. 
The results for system sizes up to $N=12$ obtained from these two post-processing schemes agree closely with each other
and show overall satisfactory  agreement with exact solutions, even without employing additional error-mitigation techniques.

Several directions merit further exploration. Finite-rate quenches, in which systems' parameters change gradually 
over time~\cite{UFischer1,UFischer2,AdiabaticXXZ,AdiabaticIsing}, offer a natural extension of the present work.
A complete characterization of DQPTs in the dimerized XXZ chain away from the flat-band limit would further deepen our understanding of
non-equilibrium dynamics in interacting topological systems. 
On the experimental side, developing shallow-circuit variants of the Hadamard test
could enable efficient amplitude estimation for intermediate system sizes~\cite{Shallow_Hq2023,Cirac,Cirac1,Shallow_H,Shallow_H1,Shallow_H2}.
More broadly, designing scalable and hardware-efficient protocols for measuring Loschmidt echoes and probing dynamical phase transitions remains a
key challenge for using intermediate-scale quantum devices in the study of complex quantum dynamics.

\begin{acknowledgments}
We are grateful to Anders Sandvik and Bal\'azs D\'ora for valuable discussions.
We acknowledge support from the IBM-Q Hub at National Taiwan University, the National Science and Technology Council (NSTC),
and the National Center for Theoretical Sciences. 
Y.C.L. was supported by the NSTC under Grants No.~NSTC 114-2119-M-007-013 and 113-2119-M-007-013; she is grateful to H.-C. Hsu and H.-C. Chang for previous collaboration. 
F.I. was supported by the Hungarian Scientific Research Fund under Grant No.~K146736 and by the National Research, Development and Innovation Office of
Hungary (NKFIH) within the Quantum Information National Laboratory of Hungary;
he also acknowledges travel support from the NSTC and National Chengchi University.     
\end{acknowledgments}

\appendix

\section{Entanglement entropy for $n=4$ XX- and XXX-chains}
\label{sec:S_n4}
In this Appendix, we provide the analytical expressions for the dynamical von Neumann EE for the $n=4$ XX- and XXX-chains, i.e. 
with $\Delta=0$ and $\Delta=1$.

For the $n=4$ XX chain, the eigenvalues of the reduced density matrix for a half-chain are given by
%
%\begin{align}
%\lambda_{1,2}=&\frac{1}{2}\left[\cos^2 \frac{Jt}{2}+\frac{1}{8}\sin^4 \frac{Jt}{2} \pm \sqrt{\cos^2 \frac{Jt}{2}+\frac{1}{4}\sin^4 \frac{Jt}{2} }\right]\;,\nonumber\\
%\lambda_{3,4}=&\frac{1}{16}\left[1-\cos^4 \frac{Jt}{2} \pm \sqrt{1-2\cos^4 \frac{Jt}{2}+\cos^8 \frac{Jt}{2}-\sin^8 \frac{Jt}{2} }\right]\;,\nonumber\\
%\lambda_5=&\frac{1}{16}\sin^4\frac{Jt}{2}\;.
%\end{align}
\begin{align}
\lambda_{1,2}=&\frac{1}{2}\left[\cos^2 \theta+\frac{1}{8}\sin^4 \theta \pm \sqrt{\cos^2 \theta+\frac{1}{4}\sin^4 \theta }\right]\;,\nonumber\\
\lambda_{3,4}=&\frac{1}{16}\left[1-\cos^4 \theta \pm \sqrt{1-2\cos^4 \theta+\cos^8 \theta-\sin^8 \theta }\right]\;,\nonumber\\
\lambda_5=&\frac{1}{16}\sin^4 \theta\;,
\end{align}
with $\theta=Jt/2$.
Here, $\lambda_{1}$ and $\lambda_{2}$ are non degenerate, $\lambda_{3}$ and $\lambda_{4}$ are both 4-fold degenerate and $\lambda_5$ is 6-fold degenerate.
From Eq.~\eqref{eq:entropy_def1}, we obtain the half-chain von Neumann EE at time $t$:
\be
{\cal S}_{1,\text{XX}}^{n=4}(t)=-4\left[\sin^2\frac{Jt}{4} \log_2 \sin^2\frac{Jt}{4} + \cos^2\frac{Jt}{4} \log_2 \cos^2 \frac{Jt}{4} \right]\;.
\label{eq:S1_XXn4}
\ee

For the XXX chain, the eigenvalues of the reduced density matrix for a half-chain are
\begin{align}
\lambda_1=&\frac{1}{16}\sin^4 \theta\;,\nonumber\\
\lambda_{2}=&\frac{1}{16}\sin^2 \theta \left(1+3\cos^2 \theta\right)\;,\nonumber\\
\lambda_{3}=&\frac{1}{16}\left(1+3\cos^2 \theta \right)^2\;,
\end{align}
where $\lambda_1$ is 9-fold degenerate, $\lambda_2$ is 6-fold degenerate and $\lambda_3$ is non-degenerate.
The half-chain von Neumann EE at time $t$ is then given by
\begin{align}
{\cal S}_{1,\text{XXX}}^{n=4}&(t)=-\frac{3}{2}\sin^2\frac{Jt}{2}\log_2 \sin^2\frac{Jt}{2} \nonumber \\
&-\frac{1}{2}(1+3 \cos^2\frac{Jt}{2}) \log_2\left(1+3 \cos^2\frac{Jt}{2}\right)+4\;.
\label{eq:S1_XXXn4}
\end{align}

\section{Shadow estimator for the twist order parameter}
\label{sec:Uz}
Consider a single qubit in the state $\rho_j$ described by the standard Pauli expansion 
$\rho_j = (\bm{I}+ \vec{r}\cdot\vec{\bm{P}})/2$ with the Bloch vector $\vec{r}$. In the measurement setting $m$ for
a fixed Pauli basis $P_j^{(m)}\in \{\bm{X}, \bm{Y}, \bm{Z} \}$, the state $\rho_j$ after a measurement is projected to
\be
        \mathcal{M}(\rho_j) = \frac{1}{2}(\bm{I} + s_j \bm{P}_j^{(m)})\,, 
\ee 
where $s_j=1-2z_j \in \{1,\, -1 \}$. Applying the inverse of the measurement channel 
$\mathcal{M}^{-1}(\,\cdot \,) = 3 (\,\cdot \,) -\bm{I}$~\cite{Shadow} and taking the tensor product over all qubits
for the $N$-qubit state, we obtain a single-shot classical shadow
\be
   \hat{\rho}^{(m)} = \bigotimes_{j=1}^N \frac{1}{2}\left( 3 s_j \bm{P}_j^{(m)} + \bm{I} \right)\,.
\ee
We then use $\hat{\rho}^{(m)}$ to construct the estimator for the twist order parameter $\e{U_z} = \sum_\mathcal{P} \alpha_\mathcal{P} \e{\bm{O}_\mathcal{P}}$. 
Using the linearity of the trace, we obtain 
\be
        \e{U_z}^{(m)} = \sum_{\mathcal{P}} \alpha_\mathcal{P} \text{Tr}[\bm{O}_\mathcal{P} \hat{\rho}^{(m)} ]\,,
\ee
where
\be
        \text{Tr}[\bm{O}_\mathcal{P} \hat{\rho}^{(m)} ] = \prod_j \text{Tr}\left [\frac{3}{2} s_j \bm{P}_{j}^{(m)} \bm{O}_j + \frac{1}{2} \bm{O}_j  \right]\,.
        \label{eq:eO}
\ee
The trace in Eq.~\eqref{eq:eO} can be simplified using the orthogonality of Pauli operators,
$\text{Tr}[\bm{P}\bm{P}']=2\delta_{\bm{P} \bm{P}'}$. 
This property implies that only qubits in $\mathcal{P}$ measured in the $\bm{Z}$ basis contribute non-vanishingly
to the trace, yielding
\be
     \text{Tr}[\bm{O}_\mathcal{P} \hat{\rho}^{(m)} ] = \begin{cases}
           3^{\abs{\mathcal{P}}} \prod_{j \in \mathcal{P}} s^{(m)}_j & \text{if } \bm{P}_{j}^{(m)} = \bm{Z}_j,\; \forall j \in \mathcal{P} \\
           0 & \text{otherwise.}
      \end{cases}
\ee
for a  single shot. We define $\eta_\mathcal{P} =  \prod_{j \in \mathcal{P}} s_j$ and average it over $N_M$ shots within 
a fixed setting $m$
\be
      \eta^{(m)}_\mathcal{P} = \frac{1}{N_M}\sum_{k=1}^{N_M} \prod_{j \in \mathcal{P}} s^{(m)}_j\,.
\ee
Finally, averaging the estimates over all $N_U$ setting, we then obtain the estimator for the twist order parameter:
\begin{align}
      \e{\hat{U}_z} & = \frac{1}{N_U} \sum_{m=1}^{N_U} \e{U_z}^{(m)} \nonumber \\
                    & = \frac{1}{N_U} \sum_{m=1}^{N_U} \sum_{\mathcal{P} \subseteq \mathcal{Z}^{(m)}} \alpha_\mathcal{P}\cdot 3^{\abs{\mathcal{P}}} \eta^{(m)}_\mathcal{P}\,,
\end{align}
with $\mathcal{Z}^{(m)} = \{j \mid \bm{P}^{(m)}_j = \bm{Z}_j \}$. 
%\be
%      \mathcal{Z}^{(m)} = \{j \mid \bm{P}^{(m)}_j = \bm{Z}_j \}\,,
%\ee

\end{document}